\documentclass{emulateapj}
\usepackage{natbib}

\shorttitle{Formation of Interstellar CO$_2$ and Other Ices}
\shortauthors{Garrod \& Pauly}

\begin{document}

\title{On the Formation of CO$_2$ and Other Interstellar Ices}

\author{R.T. Garrod}
\affil{Department of Astronomy, Cornell University, Ithaca, NY 14853,
USA} \email{rgarrod@astro.cornell.edu}

\and

\author{T. Pauly}
\affil{Iowa State University, Ames, IA 50011, USA}

\begin{abstract}

We investigate the formation and evolution of interstellar dust-grain ices under dark-cloud conditions, with a particular emphasis on CO$_2$. We use a three-phase model (gas/surface/mantle) to simulate the coupled gas--grain chemistry, allowing the distinction of the chemically-active surface from the ice layers preserved in the mantle beneath. The model includes a treatment of the competition between barrier-mediated surface reactions and thermal-hopping processes. The results show excellent agreement with the observed behavior of CO$_2$, CO and water ice in the interstellar medium. The reaction of the OH radical with CO is found to be efficient enough to account for CO$_2$ ice production in dark clouds. At low visual extinctions, with dust temperatures $>\sim$12 K, CO$_2$ is formed by direct diffusion and reaction of CO with OH; we associate the resultant CO$_2$-rich ice with the observational polar CO$_2$ signature. CH$_4$ ice is well correlated with this component. At higher extinctions, with lower dust temperatures, CO is relatively immobile and thus abundant; however, the reaction of H and O atop a CO molecule allows OH and CO to meet rapidly enough to produce a CO:CO$_2$ ratio in the range $\sim$2--4, which we associate with apolar signatures. We suggest that the observational apolar CO$_2$/CO ice signatures in dark clouds result from a strongly segregated CO:H$_2$O ice, in which CO$_2$ resides almost exclusively within the CO component. Observed visual-extinction thresholds for CO$_2$, CO and H$_2$O are well reproduced by depth-dependent models. Methanol formation is found to be strongly sensitive to dynamical timescales and dust temperatures.

\end{abstract}


\section{Introduction}

Interstellar ices are detected in both quiescent and star-forming regions in the ISM, through infrared absorption line spectroscopy.  These ices exist as mantles around a dust-grain core, and are believed to form by the accretion of gas-phase atomic and molecular material onto the grain surfaces, whereupon further chemical processing occurs. Observational data provide a number of constraints both on the chemical compositions of these ices and on their physical structure. Water ice constitutes the primary component of interstellar ice, but significant quantities of other molecules are also present, including CO, CO$_2$ and methanol (CH$_3$OH) (Chiar et al. 1995, Whittet et al. 2001, 2007), with abundances with respect to water ranging into the tens of percent, dependent on the astronomical source (Gibb et al. 2010). Comparison of observational data with laboratory spectra may also indicate the nature of the ice mixture in which the originators of the line profiles reside, i.e. whether dominated by polar water ice or by a non-polar mixture in which H$_2$O is a minor component. 

Observations indicate that water is the first ice constituent to attain abundances sufficient to be detected in the ISM; the most recently obtained threshold for H$_2$O-ice detection (using the 3 $\mu$m line) lies at a visual extinction of $3.2\pm 0.1$ (Whittet et al. 2001), indicating that the formation/deposition of this molecule occurs under conditions in which the interstellar radiation field (ISRF) is attenuated by a visual extinction of at least 1.6 (representing the extinction from the edge to the center of the cloud). The observed (edge-to-edge) visual extinction threshold for CO$_2$ of $4.3\pm 1.0$ is close to that of water ice, while CO has a significantly greater threshold of $6.7\pm 1.6$ (Whittet et al. 2007). Gas-phase formation of CO$_2$ is far less efficient than that of CO, thus grain-surface CO (which is itself primarily formed in, and accreted from, the gas phase) is believed to be the progenitor of CO$_2$ ice. The prevalence of CO$_2$ ice, and lack of CO ice, at low extinctions requires the CO-to-CO$_2$ conversion to be highly efficient. However, a significant drop in the efficiency at higher extinctions is then required to account for the detection of CO ice itself. It is not clear whether the changes in chemical behavior are caused directly by the differing visual extinctions, or whether they are related to the gas dynamics of the clouds, or to some other time-dependent effect.

The problem is further complicated by the fact that there is no clear, efficient, chemical pathway that can account for the majority of CO$_2$ ice formation. Three processes are typically cited as the most likely:

\begin{equation}
\text{HCO} + \text{O} \rightarrow \text{CO}_{2} + \text{H} \\
\end{equation}
\begin{equation}
\text{CO} + \text{OH} \rightarrow \text{CO}_{2} + \text{H} \\
\end{equation}
\begin{equation}
\text{CO} + \text{O} \rightarrow \text{CO}_{2} \\
\end{equation}

Reaction (1) is barrierless, reaction (2) is typically assigned a small activation energy barrier (80 K, Ruffle \& Herbst, 2001), and reaction (3) is thought to require a more substantial activation energy of 290 -- 1000 K (Roser et al. 2001; d'Hendecourt et al. 1985). The reactants O and OH are thought to be prevalent on the grains, as they are involved in the formation of water ice. However, the HCO radical required for reaction (1) is formed by the hydrogenation of CO on the grains, and this process has an activation energy barrier of at least 500 K (Woon, 2002). Chemical models of interstellar clouds typically find rather that it is reaction (3) that shows the most promise for CO$_2$ formation; in contrast to HCO, CO is expected to attain at least moderate abundances on the grain surface, even if CO$_2$ formation is reasonably efficient. Reaction (2) is usually found to be hampered by the relative immobility of CO and OH at low temperatures; hydrogenation of these species is far more competitive. However, Oba et al. (2010) recently conducted surface experiments to isolate reaction (2), allowing CO to react with the non-energetic products of H$_2$O photolysis, at temperatures as low as 10 K. The process was found to produce detectable quantities of CO$_2$.

Tielens \& Hagen (1982) suggested that the formation of CO$_2$ on grains requires reaction (3) to proceed with high efficiency, and later studies such as Ruffle \& Herbst (2001) and Hassel et al. (2010) have found that only models with severely reduced activation energy barriers for reaction (3) are capable of reproducing observed CO$_2$ ice abundances; atomic oxygen is otherwise too quickly hydrogenated before reaction (3) can occur. Even such low-barrier models cannot apparently reproduce the extinction dependence of CO and CO$_2$ formation.  More recently, Charnley \& Rodgers (2009) demonstrated, using a stochastic three-phase chemical model, that reaction (3) could not permit the co-existence of polar and apolar CO$_2$ ices on a single grain; these authors instead proposed reaction (1) as the primary formation mechanism.

Goumans \& Andersson (2010) recently calculated the temperature-dependent rate of reaction (3) in the gas phase. They found the rate to be yet lower than is typically assumed in models. However, they also suggested that CO and O may form a loosely-bound O...CO complex on the grain surface; the hydrogenation of the oxygen atom in the complex would immediately lead to the formation of a highly-excited OH...CO system that could easily overcome the moderate barrier to reaction (2). This mechanism has yet to be tested in computational models.

The formation of simple dust-grain ices now appears to be crucial to the formation of much more complex molecules observed in the gas phase toward star-forming regions. Ubiquitous molecules such as methanol (CH$_3$OH) and methyl formate (HCOOCH$_3$) exhibit interstellar abundances much greater than may be accounted for by gas-phase processes (e.g. Horn et al. 2004, Geppert et al. 2006, Garrod et al. 2007). Understanding the earliest phases of ice formation is therefore essential to our knowledge of the chemistry of later evolutionary stages of star formation.

Here we employ a three-phase chemical model to distinguish between the chemically-active surface layer and the deeper, inactive, mantle layers within the ice. The formation of the ice mantle allows the preservation of the chemical composition of earlier surface layers during the evolution of the cloud. By monitoring the deposition of mantle material over time, the model reveals the composition of each layer within the mantle, allowing the specific ice mixtures within which important molecules reside to be assessed, and the relationship between ice mantle composition and earlier gas-phase chemistry to be determined. In this paper we analyse the modeled abundances of the main observationally-determined components of interstellar dust-grain ices, and examine their formation mechanisms and time-dependent behavior.

In the following section we introduce the details of the chemical model and highlight changes made in the treatment of chemical processes. In section 3 we present the results of the model, applied under a variety of physical conditions. A discussion of the results follows in section 4. The conclusions are summarized in section 5.

\section{Method}

We use a gas-grain chemical model to investigate the time-dependent evolution of chemical abundances in a cold dark interstellar cloud. The basic model is that used by Garrod et al. (2008), and includes the grain-surface and gas-phase reactions used by those authors, as well those added to the network by Belloche et al. (2009). The gas-phase reaction rates not associated with the addition of new complex molecules by the aforementioned authors derive from the {\em osu.2005} chemical network. The network includes approximately 5700 gas-phase reactions, and 2900 surface/mantle processes including evaporation, accretion, and surface--mantle transport.

We employ the grain-surface binding/desorption energies ($E_{\mathrm{des}}$) and diffusion barriers ($E_{\mathrm{dif}}$) of Garrod \& Herbst (2006) and those values added by Garrod et al. (2008), all of which are representative of a surface composed primarily of amorphous water-ice. Table 1 shows a selection of values for the most important surface reactants. The diffusion barrier of each surface species, $i$, is uniformly fixed to a fraction of the associated binding energy, initially set to $E_{\mathrm{dif}}(i)=0.5 \, E_{\mathrm{des}}(i)$. Values ranging from $0.3$ to $0.8 \, E_{\mathrm{des}}(i)$ have been adopted in the past (Tielens \& Hagen 1982; Katz et al. 1999; Ruffle \& Herbst 2000; Cuppen et al. 2009); however, within this range, the values are generally very poorly constrained.

The basic $E_{\mathrm{des}}$ and $E_{\mathrm{dif}}$ values are further adjusted according to the composition of the ice surface -- in particular, the fraction of the ice composed of molecular hydrogen, H$_2$. When assuming dust temperatures $\leq 10$ K, under conditions where H$_2$ is abundant in the gas phase, gas-grain models can produce solutions in which the dust-grain ices are composed primarily of H$_2$, and the problem is exascerbated by the use of a three-phase chemical model (see section 2.2). This behavior is unphysical, and is caused by the use of binding energies that are appropriate to a water-ice surface, rather than to a mixed H$_2$:H$_2$O surface, which should demonstrate much weaker binding. Cuppen et al. (2009) estimate that the binding energy to an H$_2$ surface is around 10 times weaker than that for CO. Since, as also explained by Cuppen et al., the total binding energy is an aggregate of contributions from neighboring surface constituents, we calculate effective binding energies and diffusion barriers according to the fractional coverage of the surface with H$_2$, $\theta($H$_{2})$, giving:

\begin{equation}
E_{\mathrm{des, eff}} = E_{\mathrm{des}} \, [1-\theta(\text{H}_{2})] + 0.1 \, E_{\mathrm{des}} \, \theta(\text{H}_{2}) \\
\end{equation}

\noindent with the same relationship for $E_{\mathrm{dif, eff}}$. So long as the modifying factor used in equation (4), here 0.1, is significantly less than unity then the precise value assumed should be of small consequence. Since molecular hydrogen is the only chemical species likely to build up an appreciable grain-surface abundance that could alter the surface binding properties so strongly, as compared to the water-ice surface, we restrict these considerations only to the H$_2$ fraction. However, for completeness we allow the binding properties of all species to be altered according to this fraction, as per equation (4). This treatment renders $E_{\mathrm{des}}$ and $E_{\mathrm{dif}}$ as time-dependent quantities, making the differential equations somewhat stiffer than otherwise. A more comprehensive treatment of effective binding energies may be attempted in future. The current method produces maximum H$_2$-ice fractions of around 10\% or less, under dark-cloud conditions in which molecular hydrogen dominates in the gas phase. The variation of the binding energies tends to flatten out quickly as the H$_2$ fraction stabilizes, according to the balance between accretion and thermal evaporation; other processes have a minimal effect on surface H$_2$ abundance.

The grain-surface chemistry is treated using the modified-rate method of Garrod (2008), to account for stochastic effects where present. The method has been shown by Garrod et al. (2009) to produce an excellent match to exact Monte Carlo simulation methods, over a broad range of physical parameters. A canonical grain size of $0.1 \mu$m is assumed, with $10^6$ surface binding sites.

The model includes gas-phase and grain-surface cosmic-ray photodissociation/photoionization processes (as per Garrod et al. 2008), assuming a cosmic-ray ionization rate of $\zeta = 1.3 \times 10^{-17}$ s$^{-1}$. The initial elemental abundances used in all models are shown in table 2, and are the same as those used by Garrod et al. (2008).

We include the reactive desorption mechanism given by Garrod et al. (2007), whereby the formation of a molecule on the grain surface can result in its desorption into the gas phase, with a probability of <1 \%. We assign a value $a_{\mathrm{RRK}}=0.01$ for this process. We also include photodesorption for all surface species, using the rates and yields of {\"O}berg et al. (2009a,b); yields of 10$^{-3}$ are assumed for species for which no data currently exist. For visual extinctions much larger than 1, reactive desorption strongly dominates over direct photodesorption; although, in some cases, the reactive desorption process is preceded by the photodissociation of a surface molecule by the cosmic ray-induced UV field.

\subsection{Reaction--diffusion competition}

The efficiencies of activation-energy barrier-mediated reactions on the grain surfaces are crucial to this study. In general, under the classical rate-equation expression, the rate of a reaction A+B is expressed as:

\begin{equation}
k_{AB}=f_{act}(AB) \times [k_{hop}(A)+k_{hop}(B)]/N_{S} \\
\end{equation}

\noindent where $k_{hop}$ represents the rate at which each reactant migrates from one surface binding site to another, by a process of thermal hopping,  $N_S$ is the number of binding sites on the surface, and $f_{act}(AB)$ is an efficiency factor related to the activation energy barrier. This reaction rate represents the rate at which two reacting species migrate over the surface of a grain and meet in the same binding site, and there react. The efficiency $f_{act}(AB)$ is usually assigned a value equal to the probability of reaction upon a single collision between the reactants, $\kappa(AB)$, expressed as a Boltzmann factor, $\kappa(AB)=exp(-E_{A}/T)$, or a more complex expression for quantum tunneling (see section 2.2). However, in reality, a reaction may have many more opportunities to occur, because the reactants are confined in the same binding site until one or other reactant hops into an adjacent site (Tielens \& Hagen 1982). Thus, in cases where activation energy barriers exist, the competition between reaction and migration must be considered. We use an expression similar to that given by Chang et al. (2007) (different only in that we ignore evaporation, as its rate is negligible compared to thermal hopping):

\begin{equation}
f_{act}= \frac{\nu \kappa(AB)}{\nu \kappa(AB) + k_{hop}(A) + k_{hop}(B) } \\
\end{equation}

\noindent where $\nu$ is the frequency of collision of the two reactants. Here we take $\nu$ to be equal to the larger of the characteristic vibrational frequencies of the two reactants, as calculated by the method given by Hasegawa et al. (1992) and typically on the order of $10^{12}$ Hz. The result of equation (6) is that in the limit where thermal hopping is much faster than the barrier-mediated reaction rate, $\nu \kappa(AB)$, the overall rate of reaction, $k_{AB}$, depends only on the reaction probability, $\kappa(AB)$, and not on the hopping rates. Conversely, if the reaction probability is sufficiently high, the overall reaction rate is limited only by the rate of hopping.

\subsection{Three-phase model}

To the basic chemical model, which consists of a coupled gas-phase and grain-surface chemistry, we add a third phase that represents the ice mantle that forms beneath the outer ice surface. This three-phase model (gas/surface/mantle) is based on the formulation presented by Hasegawa \& Herbst (1993). In this formulation, the surface layer represents the outermost layer of ice, which resides either on top of other ice layers, or on the surface of the grain itself. The surface and gas-phase chemistry are coupled through the accretion of material from the gas phase and the evaporation and non-thermal desorption of material from the surface back into the gas phase. As the surface coverage of the grain increases, the mantle is allowed to grow, incorporating material from the surface. The rate of growth of the ice mantle is determined by the net rate of deposition of material onto the grain (which takes into account thermal and non-thermal desorption {\em from} the grain surface), and by the degree of surface coverage of the bare grain. Thus, the mantle begins to form even before full surface coverage of the grain is achieved, representing deposition of new material on top of other surface-bound material. When full surface coverage is achieved, any net gain in surface material results in a commensurate transfer of material to the mantle, with surface material replaced by new material from the gas phase. This transfer does not, of course, represent a physical motion of material between surface and mantle; the transfer is rather an accounting issue within the chemical model, as the surface layer is continually renewed.

Following Hasegawa \& Herbst (1993), we define the surface concentration of species $i$ as $n_{s}(i) = N_{s}(i) \, n_{dust}$, where $N_{s}(i)$ is the average, absolute number of particles of species $i$ present in the surface layer and $n_{dust}$ is the gas-phase concentration of dust particles. Thus, the quantity $N_{S} \, n_{dust}$ represents one monolayer of ice in this parameterization. The calculation of the net loss or gain of each species due to chemical and physical processes (specifically excluding any loss or gain due to transfer between surface and mantle), $[dn_{s}(i)/dt]_{chem}$, is achieved using the rates determined in the normal operation of the chemical code. The net rate of change in total surface material is thus $[dn_{s}/dt]_{chem}=\sum_{i}[dn_{s}(i)/dt]_{chem}$. This quantity is used to calculate both the total rate of transfer from surface to mantle, and the rate of transfer for individual surface species, respectively,

\begin{equation}
[dn_{s}/dt]_{tran} = \alpha_{acc} [dn_{s}/dt]_{chem} \\
\end{equation}

\begin{equation}
[dn_{s}(i)/dt]_{tran} = [dn_{s}/dt]_{tran} [n_{s}(i)/n_{s}] \\
\end{equation}

\noindent where $n_{s}=\sum_{i}n_{s}(i)$, and $\alpha_{acc}$ is the fractional surface coverage. The resultant loss/gain rates for the transfer of surface/mantle species are appended to the chemical rates to give the total loss/gain rates,

\begin{equation}
[dn_{s}(i)/dt]_{tot} = [dn_{s}(i)/dt]_{chem} - [dn_{s}(i)/dt]_{tran} \\
\end{equation}

\begin{equation}
[dn_{m}(i)/dt]_{tot} = [dn_{m}(i)/dt]_{chem} + [dn_{s}(i)/dt]_{tran} \\
\end{equation}

\noindent where $n_{m}(i)$ represents the concentration of mantle species $i$.

In the case where $ [dn_{s}/dt]_{chem} < 0$, and thus there is net desorption of material from the ice surface, new material is transferred from the mantle to the surface, according to the expressions

\begin{equation}
[dn_{s}/dt]_{tran} = \alpha_{des} [dn_{s}/dt]_{chem} \\
\end{equation}

\begin{equation}
[dn_{s}(i)/dt]_{tran} = [dn_{s}/dt]_{tran} [n_{m}(i)/n_{m}] \\
\end{equation}

Contrary to Hasegawa \& Herbst (1993), we set $\alpha_{des}=n_{m}/n{s} \neq \alpha_{acc}$, with $\alpha_{des}$ limited to a maximum value of unity. This change allows all surface material to be available for desorption, whereas the adoption of  $\alpha_{des}=\alpha_{acc}$ requires the complete removal of the mantle before the degree of surface coverage is allowed to fall to less than 1.

In this model we assume the surface material, of a thickness of 1 monolayer, to be chemically active, while the mantle is taken to be entirely inert. At the grain temperatures of around 10 K that are applied to models in this study, the mobility of mantle-bound species is likely to be negligible. Furthermore, the ability of surface hydrogen atoms to penetrate the surface and react with material in the mantle layers has been shown by Fuchs et al. (2009) also to be negligible at such low temperatures, based on microscopic models of laboratory ice experiments. While this treatment is applicable to ices detected in quiescent regions, the further chemical processing of layers deep within the ice mantle engendered by the star-formation process would require a more complex chemical treatment of the ice mantle.

\subsection{Activation-Energy Barriers and Quantum Tunneling}

It is well known that a number of important grain-surface reactions are mediated by activation energy barriers. The grain-surface formation of CO$_2$, as well as that of formaldehyde (H$_2$CO) and methanol (CH$_3$OH), appears to require reactions with significant activation-energy barriers. The O + CO $\rightarrow$ CO$_2$ reaction is freqently invoked as a primary formation route for carbon dioxide, due to the availability of the reactants; however, it is mediated by a barrier, most recently estimated by Roser et al. (2001) to be of height 290 K.

The activation energies of many pertinent reactions have been ascertained with at least moderate precision through experiment or calculation. However, a knowledge of the activation energy alone provides only a very weak constraint on the rate of reaction at extremely low temperatures. This is because the reaction probabilities at around 10 K are so small that quantum tunneling processes can become the dominant means by which a reaction proceeds, especially for reactions involving atomic hydrogen, and to some degree other atomic species such as C and O (see Goumans \& Andersson 2010). The rates of such quantum processes are generally very poorly constrained. Chemical models such as that of Hasegawa et al. (1992) have parametrized the tunneling process by adopting an expression for tunneling through a rectangular potential:

\begin{equation}
\kappa = \exp[-2(a / \hbar) (2 \mu E_{A})^{1/2}] \\
\end{equation}

\noindent where $\mu$ is the reduced mass, and $a$ is the width of the barrier, assigned a uniform value of 1 {\AA} by Hasegawa et al. (1992). Subsequent models have typically taken the same approach (e.g. Garrod et al. 2007); however, the assumed barrier width is arbitrary.

Goumans \& Andersson (2010) recently calculated the temperature-dependent rate of reaction (3) in the gas phase, using quantum transition state theory. These calculations imply a reaction probability per collision at 10 K of $\kappa=4.8 \times 10^{-23}$ (T. Goumans, private comm.). This probability represents a purely quantum tunneling process, as the thermal rate is negligible at low temperatures. Parametrizing this value using equation (13), and assuming $E_{A}=290$ K (Roser et al. 2001), suggests a barrier width of 2.33 {\AA}.

Andersson et al. (2009) calculated gas-phase rates for the hydrogen abstraction reaction H + CH$_4$ $\rightarrow$ CH$_3$ + H$_2$. That study suggests a reaction probability per collision at 10 K of  $\kappa=3 \times 10^{-30}$ (T. Goumans, private comm.). This value is around 15 orders of magnitude less than that obtained using equation (13) with a barrier width of 1 {\AA} and a height of $E_{A}=5940$ K, as assumed in previous models. Assuming the same activation energy, the Andersson et al. value  suggests a square-well potential width of 2.17 {\AA}.

Fuchs et al. (2009) conducted laboratory experiments on the hydrogenation of CO ice, and modeled this system using microscopically exact methods that naturally take account of reaction--diffusion competition. They found that the H+CO and H+H$_2$CO hydrogenation reactions proceed by a tunneling process; their combined models and data imply a reaction probability of approximately $\kappa=10^{-14}$, to within around 1 order of magnitude. 

Here we adopt the 10 K tunneling probabilities derived from the results of Goumans \& Andersson (2010) and Andersson et al. (2009). For the grain-surface hydrogenation of CO and H$_2$CO, and in the absence of directly measured or calculated rates, we assume the typical activation-barrier heights of 2500 K (e.g. Garrod et al. 2008), but use a barrier width $a=2$ {\AA} to calculate the square-well tunneling rates according to equation (13).

\subsection{Treatment of the CO+OH$\rightarrow$CO$_2$+H reaction}

The precise mechanism for reaction (2) involves multiple transition states, whose relative energies are somewhat uncertain (e.g. Goumans et al. 2008, Song et al. 2006, Frost et al. 1993). However, a simplistic description of the mechanism is that CO and OH must overcome a small activation-energy barrier to form an excited intermediate, HOCO*. The exit channel to form CO$_2$ + H from HOCO* is of a similar barrier height to that of the reverse entry channel that returns CO + OH. However, it is likely that quantum tunneling through this barrier makes the formation of CO$_2$ the most probable outcome (Frost et al. 1993, Chen \& Marcus 2005), if collisional de-excitation of HOCO* does not occur first. Goumans et al. (2008) suggest that such de-excitation may be possible on dust-grain surfaces, through coupling to the ice surface, at a rate sufficient to make stable HOCO the most likely product. These authors also aver that under interstellar conditions, reaction of this product with atomic hydrogen could lead to a selection of products, including CO$_2$ and formic acid, HCOOH. The large abundance of interstellar atomic hydrogen, and its high mobility on grain surfaces, would make such a mechanism highly efficient. In such a case, the rate limiting step in the formation of CO$_2$ (from CO and OH already present in a surface binding site) would thus be the traversing of the initial barrier to form HOCO*. However, recent experiments by Oba et al. (2010), in which reaction (2) was isolated, have indeed shown the CO + OH reaction to be capable of forming CO$_2$ on ice surfaces at temperatures as low as 10 K, while observing no signatures from the other suggested product branches.

Previous chemical models (e.g. Ruffle \& Herbst, 2001) have assumed an activation energy barrier for reaction (2) of 80 K. This appears to be based on the measured rate of the gas-phase reaction (Frost et al., 1993), compared with the collisional rate, giving an efficiency of $\sim$10$^{-4}$. Here we adopt the 80 K value, at which the thermal reaction probability exceeds the quantum tunneling value obtained through equation (13). In fact, the consideration of competition between reaction and diffusion described in equation (6) results in a reaction efficiency $f_{act}$ that is close to unity in this case, due to the relatively high diffusion barrier for CO of 575 K (assuming $E_{\mathrm{dif}}=0.5E_{\mathrm{des}}$). Thus, in this model, reaction (2) is essentially unaffected by the small activation energy barrier, and proceeds at very close to the diffusion rate. Even adopting the highest values obtained from various calculations of the entry channel barrier (e.g. Chen \& Marcus, 2005), around 400 -- 500 K, reaction would probably dominate over migration of CO from the binding site, especially if quantum tunneling is efficient.

As well as the production of CO$_2$ by the addition of mobile CO to OH, we also incorporate into the model the possibility that an oxygen atom may be hydrogenated while situated on top of a surface CO molecule. The resultant chemical energy released would be quite sufficient to overcome the small activation barrier to reaction (2), and could improve the likelihood of a rapid dissociation of HOCO* to CO$_2$ + H. We formulate this mechanism as:

\begin{equation}
\text{H} + \text{O} + \text{CO} \rightarrow \text{CO}_{2} + \text{H} \\
\end{equation}

A similar such mechanism was suggested by Goumans \& Andersson (2010), based on their calculated gas-phase O--CO potential-energy surface, in which the oxygen atom may form a loosely bound O...CO complex, allowing it to remain with the CO molecule for longer, prior to the arrival of an H atom. In the case where atomic hydrogen is reasonably abundant, such as under typical dark-cloud conditions, it is likely that the timescale for reaction of H and O would be shorter than that of thermal diffusion of an O atom. In such a case, the additional binding to the CO molecule would be irrelevant, and the distribution of O with respect to CO would be essentially random. Under this assumption, we allow that a fraction, $\theta($CO$)$, of all H + O $\rightarrow$ OH reactions leads to the formation of CO$_2$ + H, where $\theta($CO$)$ is the fractional surface coverage of CO. This formulation thus represents a minimum reaction efficiency for this mechanism, which would move (from a value $\theta($CO$)$) to something closer to unity under hydrogen-poor conditions.

\section{Results}

\subsection{Dark cloud ice chemistry}

The method described in section 2 is applied to a static single-point dark-cloud model, with density $n_{\mathrm{H}}=2 \times 10^{4}$ cm$^{-3}$ and visual extinction $A_{\mathrm{V}}=10$. Both gas and grains are assigned temperatures of 10 K. Figure 1 shows the growth of the ice mantle over time, up to the final model time of $10^7$ yr. The deposition of the ice is not a simple function of time; however, the rate of increase over the first 100 monolayers is approximately linear. An ice thickness of 1 monolayer is achieved after a few thousand years. The rate of mantle deposition falls after 100 monolayers, as gas-phase material becomes depleted. Approximately 300 monolayers have formed after $10^7$ yr, by which point the (net) deposition rate has fallen close to zero.

Figure 2(a) shows the chemical composition within each layer of the ice. A timescale is also shown along the top of the plot, which is mapped to the values of the lower x-axis according to the relationship shown in figure 1. The time corresponding to any particular ice layer represents the time in the cloud's chemical evolution at which the ice layer was deposited. The chemical composition of the ice beneath the surface layer is preserved as the cloud evolves. Thus, the curves to the left of any chosen time in the plot represent the instantaneous composition of a section through the ice at that time.

It may be seen that water is the dominant ice constituent throughout the model run, and has a relatively constant abundance throughout the ice mantle. As found by Cuppen \& Herbst (2007), the reaction H$_2$ + OH $\rightarrow$ H$_2$O ($E_{A}=2100$ K, Baulch et al., 1984) is the main mechanism for the conversion of OH to water, rather than reaction with atomic H, because of the large H$_2$ abundance on the grains ($\sim$3 \% of surface coverage). Methane (CH$_4$) and ammonia (NH$_3$) are both abundant in the first $\sim$50 monolayers of ice, due to the atomic initial conditions; C and N are successively hydrogenated on the grain surfaces. 

Figure 2(b) shows the gas-phase abundances (with respect to total hydrogen) of a selection of species, also mapped to the ice deposition profile for ease of comparison. NH$_3$ ice abundance gently falls, through the outer layers, as the abundance of CO and its products take up a greater proportion of the ice composition. The dramatic fall in the abundance of CH$_4$ ice is caused by the steady removal of atomic C from the gas phase, caused primarily by its conversion into gas-phase CO.  Gas-phase atomic oxygen is initially much less affected by its conversion to CO, due to its larger abundance. As more CO is formed in the gas phase, more is accreted onto the grains, and CO reaches $\sim$10 \% of the total ice ($\sim$20 \% with respect to water) between monolayers 100 -- 250. Gas-phase CO reaches its peak abundance at around $3 \times 10^{5}$ yr, but it only grows to dominate over all species (excluding H$_2$) by around $7 \times 10^{5}$ yr.

CO$_2$ ice is formed in appreciable abundances throughout the model run, typically 2 -- 4 times less abundant than CO. Both CO and CO$_2$ ice abundances follow the gas-phase behavior of CO. The only active grain-surface CO$_2$ formation route is that of equation (14), in which the H + O $\rightarrow$ OH reaction occurs on top of a CO molecule, leading to the immediate formation of CO$_2$ + H. If this mechanism is turned off in the model, CO$_2$ abundances reach little more than a fraction $10^{-3}$ of CO abundance; in that case, reaction (3) is the dominant production mechanism.

The gas-phase abundance of atomic hydrogen stays fairly constant at $\sim$$10^{-4} n_{\mathrm{H}}$ throughout the model run, even as heavier elements deplete from the gas phase. However, as the rate of accretion of atomic oxygen from the gas-phase falls, the ability of surface oxygen to remove hydrogen atoms on the grains (by the formation of OH and H$_2$O) diminishes, and atomic H abundance on the grains increases. The fall in water ice production also results in a slowing of the rate of ice deposition at late times ($>10^{6}$ yr), as seen in figure 1. This reduces the rate at which CO (and other surface species) are overlaid with new material, leaving them more vulnerable to hydrogenation by accreting H atoms. This combined effect of the rise in grain-surface atomic hydrogen abundance and the slower rate of ice deposition leads to a much greater efficiency in the hydrogenation of CO and H$_2$CO to methanol (CH$_3$OH). The effect is further enhanced as CO takes over as the dominant accreting species other than hydrogen, at $\sim$1 Myr. Thus, the formation of large quantities of methanol in the outer ice mantle layers is a symptom of the depletion of gas-phase oxygen and its conversion in the gas-phase into CO, under conditions where atomic H is abundant. Charnley \& Rodgers (2009) found a similar effect in their model, albeit apparently influenced by the action of hydrogen abstraction reactions on surface methanol and its precursors.

Figure 2(b) also shows the abundance of {\em gas-phase} methanol, as a function of the number of ice layers in the mantle. The CH$_3$OH fractional abundance matches the canonical dark-cloud value of $\sim$$10^{-9}$, from around $10^{5}$ yr until near to the end of the run. This methanol is formed on the dust-grain surfaces, and is ejected into the gas phase with an efficiency of $\lesssim 1$\% by the chemical energy released upon its formation, under the formulation given by Garrod et al. (2007). At the 10 magnitudes of visual extinction used in this model, direct photodesorption at the yields determined by {\"O}berg et al. (2009a,b) is found to be negligible compared to reactive desorption. While the conversion of CO to CH$_3$OH on the grains becomes more efficient at later times, the overall rate of conversion remains fairly steady, due to the falling rate of deposition of CO onto the grains from the gas phase. This leaves the gas-phase abundance of methanol stable over a long period.

Molecular oxygen is formed in the gas-phase, mainly by the reaction O + OH $\rightarrow$ O$_2$ + H; molecular oxygen abundance is shown in figure 2(b). This O$_2$ can accrete onto the grain surface and contribute to the formation of surface OH and water. This occurs primarily by the sequential hydrogenation of O$_2$ by atomic H, to produce hydrogen peroxide, H$_2$O$_2$, which is followed by the hydrogen abstraction reaction H + H$_2$O$_2$ $\rightarrow$ H$_2$O + OH. The reactions of H with O$_2$ and H$_2$O$_2$ are assigned activation energy barriers of 1200 K and 1400 K respectively (Tielens \& Hagen 1982), following the networks of Ruffle \& Herbst (2001) and earlier models by those authors. However, the treatment of competition between reaction and diffusion used in the present model (see section 2.1), combined with quantum tunneling through the activation barriers (section 2.3), allows both reactions to proceed at close to the hydrogen diffusion rate. This high efficiency is in line with experimental findings by Miyauchi et al. (2008), Ioppolo et al. (2008), and Oba et al. (2009) showing that O$_2$ may be efficiently converted to water at low temperatures. In this model, the O$_2$ to H$_2$O conversion is sufficiently rapid that there is no significant build-up of O$_2$ (or indeed H$_2$O$_2$) in the ice mantles. The contribution of O$_2$-related chemistry to the water-ice formation rate is around 1 \% until close to the end of the model run, when gas-phase O$_2$ briefly dominates over atomic O. At its peak, the H + H$_2$O$_2$ $\rightarrow$ H$_2$O + OH reaction contributes around 30 \% of water production, and $\sim 50$ \% of OH production. The major O$_2$ contribution to water formation is thus confined to the outermost ice layers, and has no discernible effect on the water profile through the ice, as shown by figure 2(a).

\subsubsection{Variations in physical/chemical parameters}

Here we test the effect on the basic dark-cloud chemical model of variations in the dust temperature, the cloud density, and the generic diffusion-barrier/binding-energy ratio.

Figure 3 shows the ice mantle composition for model runs with dust temperatures of 8 K and 12 K. The gas temperature is held constant at 10 K, in order to isolate the surface/ice-specific effects. In fact, the general behavior of the chemistry in the 8 -- 12 K range is quite robust. The main difference between models, as judged by the observationally-detectable, saturated species shown in the figure, is that the efficiency of methanol production is generally lower at higher temperatures. This is due to the greater mobility of hydrogen, which increases its reaction rate with surface radicals and thus lowers its surface abundance, while leaving unaffected the total reaction rate for barrier-mediated reactions such as H + CO $\rightarrow$ HCO and H + H$_2$CO $\rightarrow$ CH$_3$O/CH$_2$OH. Because both the formation and destruction of formaldehyde (H$_2$CO) are governed by these reactions, the abundance of H$_2$CO relative to CO is less affected by temperature changes, and broadly follows the behavior of CO.

The constancy of general behavior over the 8 -- 12 K range shown in figure 3 is largely due to the system being in the accretion limit. This means that surface reactions (without activation energy barriers) are rapid compared to the rate of accretion, and thus production occurs at these limiting accretion rates rather than according to the temperature-dependent surface-diffusion rates. However, the averaged abundances of intermediate species such as OH or CH$_3$ are strongly affected by the temperature variation; their small surface abundances (i.e. typically less than one particle per grain at any instant) reflect their lifetimes on the grain surface before reaction occurs. Reactions with activation energy barriers (which are typically much slower than accretion rates) are also largely unaffected by temperature changes, due to the dependence of their rates primarily on the activation barrier tunneling rate, as per equation (6). The variation with temperature of the overall number of ice layers formed is caused by the larger proportion of H$_2$ being incorporated into the ice mantles at lower temperatures, due to the lower rate of evaporation.

Figure 4(a) shows the results of the model using a generic assignment of $E_{\mathrm{dif}}=0.3E_{\mathrm{des}}$ (as adopted by Hasegawa et al. 1992, based on estimates given by Tielens \& Allamandola 1987) with a dust temperature of 8 K. Again, the change in the general chemical behavior is small; the lowering of diffusion barrier heights is analogous to an increase in dust temperature, in the regime modeled here. The main difference in behavior compared with the $E_{\mathrm{dif}}=0.5E_{\mathrm{des}}$ case, is the initially higher methanol ice abundance, which is unrelated to CO surface abundance. This methanol formation is enabled by the increased mobility of oxygen atoms, which react at a sufficiently fast rate with methyl (CH$_3$) radicals to contribute significantly to CH$_3$O formation, and thus to methanol production. A similar increase in surface mobilities is found in the basic model with a dust temperature of 12 K, as seen in figure 3(b); however, in that case, the abundance of CH$_3$ is lower due to the much lower molecular hydrogen abundance on the grains, whose presence otherwise provides a mechanism for the hydrogenation of C to CH$_2$ and of CH$_2$ to CH$_3$. This alternative methanol-production route falls away as the rate of accretion of atomic carbon onto the grains diminishes. 

Figure 4(b) shows the same model ($E_{\mathrm{dif}}=0.3E_{\mathrm{des}}$), except with an increased dust temperature of 12 K. In this case, the chemical behavior on the grains is far different; CO$_2$ dominates CO throughout the mantle, and even becomes the dominant ice-mantle constituent through most of the ice layers. In this case, CO has become sufficiently mobile for reaction (2) to compete with H$_2$ for reaction with OH; the elevated temperature also decreases H$_2$ coverage, which stands at less than 0.01 \% at 12 K. 

Clearly, under certain temperature conditions, or with somewhat higher mobilities of surface species, CO$_2$ can be seen to dominate the ice composition. However, at no point in the model shown in figure 4(b) does CO take over from CO$_2$; the use of a single-point static chemical model appears to be insufficient to obtain a strongly time-dependent relationship between CO and CO$_2$ abundance in the ice.

Figure 5 shows ice-mantle and gas-phase abundances using the basic $E_{\mathrm{dif}}=0.5E_{\mathrm{des}}$ model with a density reduced by a factor of ten, to $n_{\mathrm{H}}=2 \times 10^{3}$ cm$^{-3}$. Here both CO and CO$_2$ comprise only a few percent of the total ice mantle, while hydrogenation of CO to methanol is highly efficient. This is caused by the ten times higher atomic-hydrogen abundance that results from the ten times slower conversion of H to H$_2$. Here, accretion of atomic H onto the grains is much faster than that of atomic O, thus there is little competition for hydrogen between O and CO, and surface CO is rapidly converted to methanol at all times. In this low density case, the depletion of heavy elements does not reach completion by 10$^7$ yr. The somewhat lower rate of mantle formation also allows more time for CO hydrogenation before the CO molecules are covered by new ice layers.

Figure 6 shows ice-mantle and gas-phase abundances, assuming a density $n_{\mathrm{H}}=2 \times 10^{5}$ cm$^{-3}$. Here, the general behavior of the chemistry is more similar to the $n_{\mathrm{H}}=2 \times 10^{4}$ cm$^{-3}$ case, with the exception of methanol. Atomic hydrogen abundance in the gas phase is typically less than that of heavier elements or CO. Thus, oxygen reacts with most of the surface H, leaving the majority of CO intact. Only at very late times -- in the outermost ice layers -- does methanol become abundant, as heavier species become depleted and atomic hydrogen again dominates the accretion from the gas phase. Gas-phase methanol, which originates on the grains, never reaches the typical observed value of $\sim$$10^{-9}n_{\mathrm{H}}$. Total depletion of most gas-phase species occurs at around 10$^6$ yr.

\subsubsection{Discussion of dark cloud model results}

The static dark-cloud models show that a significant fraction of CO may be converted to CO$_2$ on the dust grains at low temperatures. This behavior appears quite robust over a moderate range of dust temperatures and diffusion barrier heights. The mechanism at work is reaction (2); however, the means by which it occurs is not through the diffusion of the reactants CO and OH, but rather the formation of OH {\em in situ} in a binding site that already contains a CO molecule. Thus, the only surface diffusion necessary for the formation of CO$_2$ in this case is atomic hydrogen; atomic O may simply remain where it lands following accretion. The proportion of CO that is converted to CO$_2$ is then simply dependent on the fraction of the surface composed of CO. 

The suggestion by Goumans \& Andersson (2010), that O and CO may also form a loose bond, could increase the conversion rate, where the atomic hydrogen accretion rate is low enough to allow atomic oxygen time to diffuse significantly around the grain surface. We also tested this mechanism using the basic 10 K dark cloud model, adding a new species, O...CO, to the model to represent this loosely bound association of O and CO, and allowing reaction with a selection of prevalent reaction partners including atomic H. However, we found that this approach yielded CO$_2$ production efficiencies lower than those obtained with the method outlined in section 2.4. This was due to the lack of an Eley-Rideal mechanism for the formation of O...CO, whereby oxygen could accrete directly onto a CO molecule -- the formation of O...CO in the model was thus solely dependent on the diffusion of atomic O. In the case where such an Eley-Rideal mechanism is the dominant O...CO production route, the result should reduce to that of the simpler approach in any case. The incorporation of O...CO into the reaction network in a sufficiently comprehensive fashion, including an appropriate treatment under the modified-rate method of Garrod (2008), is left for future exploration.

At somewhat higher temperatures and/or with lower diffusion barriers, the efficiency of CO to CO$_2$ conversion undergoes an enormous increase, as CO becomes sufficiently mobile on the grain surfaces to compete for reaction with OH -- reaction (2) becomes active through direct surface diffusion. It is therefore plausible that higher temperatures during the early formation of the ice mantles could produce the highly efficient CO$_2$ formation observed at low visual extinctions.

The ice composition produced by the models is in fair agreement with observational values. Table 3 shows the ice observations, as collated by Gibb et al. (2000), toward the field star Elias 16, which is commonly taken as a representative dark cloud source. Alongside, we give the calculated composition obtained with the basic 10 K model, and summed over all the ice layers present at a given time, for a range of model times. Of the times shown, $10^{6}$ yr might be chosen as the optimum; however, the abundances of CH$_4$ and NH$_3$ are excessively high at all times in the model. Judging by the results shown in figure 4, this could perhaps be remedied somewhat by the choice of a lower dust temperature and diffusion barrier height. Such changes would have no significant effect on the agreement of CO/CO$_2$ abundances, as seen in figure 4(a). The higher temperature (12 K) case seen in figure 4(b) demonstrates an improved CH$_4$ agreement, but NH$_3$ agreement is generally worse, while the CO/CO$_2$ behavior shows strong CO$_2$ dominance as mentioned above, which does not fit with observational results. However, since both CH$_4$ and NH$_3$ are particularly abundant early in the model cloud's evolution, it is also likely that a better dynamical treatment of the cloud's evolution at early times would improve the fidelity of the model to the observed values. Both CH$_4$ and NH$_3$ abundances within the ice are lower (i.e. better), under the 12 K regime, at early times (or for the deepest ice layers), while lower temperatures show better agreement at intermediate times (middle ice layers). This would suggest that the collapse and cooling of the cloud may play a significant role in controlling NH$_3$ and CH$_4$ abundances in the ice.

Methanol and formaldehyde abundances are also a little high at later times, comparing the basic 10 K model with the observational values, but variation from the assumed cloud density could change this.

\subsection{Depth-dependent models}

Here we attempt to model the visual extinction-dependent chemistry of water, CO and CO$_2$ ices, in order to address the observed extinction-threshold behavior of these species. We assume a constant density throughout a dark cloud, and apply the model at a selection of depth points from the edge of the cloud toward the center, corresponding to fixed visual extinctions.
Crucially, we adopt a visual extinction-dependent dust temperature, assuming exposure to the standard interstellar radiation field (ISRF). To calculate this temperature in a simple manner, we assume that the rate of cooling of the dust is equal to the rate of radiative heating by the ISRF, thus we solve:

\begin{equation}
\int_{0}^{\infty} \! Q_{\nu}B_{\nu}(T_{\mathrm{dust}}) \, \mathrm{d}\nu = \int_{0}^{\infty} \! Q_{\nu}J_{\nu}D_{\nu}(A_{\mathrm{V}}) \, \mathrm{d}\nu
\end{equation}

\noindent where $Q_{\nu}$ is the frequency-dependent efficiency of emission and absorption of the grain, $B_{\nu}(T_{\mathrm{dust}})$ is the Planck function for dust-grain temperature $T_{\mathrm{dust}}$, $J_{\nu}$ is the radiation field intensity, and $D_{\nu}(A_{\mathrm{V}})$ is the attenuation of the field at given frequency by the dust column ahead of the position in question. To simplify this calculation, we evaluate the left-hand side using the approximation of Kr{\"u}gel (2008), L.H.S. $\simeq 1.47 \times 10^{-6} \, a \, T_{\mathrm{dust}}^{6}$, expressed in cgs units. To evaluate the right-hand side, we use the radiation field of Zucconi et al. (2001), which consists of multiple, modified black-bodies ranging from the optical--NIR down to cosmic microwave background radiation. To evaluate the emission efficiency we use $Q_{\nu} \simeq 10^{-23} \, a \nu^{2}$ (Kr{\"u}gel 2008), with $\nu$ in Hz and $a$ in cm. We follow the approach given by Cuppen et al. (2006) to evaluate the attenuation of the radiation field, 

\begin{equation}
D_{\nu} = \exp \left( -0.8 \left[ \frac{A_{\nu}}{A_{\mathrm{V}}} \right] A_{\mathrm{V}} \right)
\end{equation}

\noindent using the tabulated values given by Mathis et al. (1990) to evaluate $A_{\nu}/A_{\mathrm{V}}$. Our treatment implicitly assumes plane-parallel geometry. Figure 7 shows the resulting dust temperature profile, using the canonical 0.1 $\mu m$ grain size; $T_{\mathrm{dust}}=8.41$ K at $A_{\mathrm{V}}=10$.

At low visual extinctions, the photodissociation rates of H$_2$ and CO become important to the chemistry. We use the tabulated values of Lee et al. (1998), which depend on visual extinction and on the column densities of H$_2$ and CO. Beginning at the edge of the cloud, the code works inward, summing the column densities through previously calculated points (at a given time step), to give the current column density values. The final ratio of H:H$_2$ for each depth point is used as the starting ratio for the next point inward. The model was run over a visual extinction range of 0 -- 10 magnitudes, from cloud edge to center. A linear relationship between total hydrogen column density and visual extinction is assumed, with $N($H$+2$H$_{2})=1.6 \times 10^{21} \, A_{\mathrm{V}}$. We adopt a fixed gas temperature of 10 K. 

Figure 8 shows the depth-dependent column-density behavior of H$_2$O, CO$_2$ and CO ices for three constant gas densities: $n_{\mathrm{H}}=2 \times 10^{3}$ (low), $6 \times 10^{3}$ (intermediate), $2 \times 10^{4}$ cm$^{-3}$ (high), assuming a generic grain-surface diffusion-barrier height $E_{\mathrm{dif}}=0.3E_{\mathrm{des}}$. Curves for each molecule are shown for three different times in the models: 0.5, 1 and 2 Myr; higher curves correspond to later times. The visual extinction values shown in the figure are twice those used in the chemistry model, so as to represent edge-to-edge values. The column density at any given visual extinction represents the sum of the calculated abundances, from the edge of the cloud inward to that point, multiplied by 2 to give an edge-to-edge value. Thus, the plotted column densities and visual extinctions are directly comparable to the observed quantities. The dotted line represents an order-of-magnitude estimate of the detection threshold for all species. More precise estimates may be found in the literature (as cited in the Introduction).

It is immediately seen that the general, observed visual-extinction threshold behavior is reproduced by the models; water ice is formed at the lowest visual extinction of each of the three molecules, with CO$_2$ reaching the threshold ($10^{16}$ cm$^{-2}$) at marginally higher $A_{\mathrm{V}}$, and CO following at significantly greater visual extinctions. The threshold for H$_2$O in the low- and intermediate-density runs is a remarkably good match to the observational value, with values of 4 and 3 magnitudes, respectively. The water threshold is fairly insensitive to the age of the cloud, over the 0.5 -- 2 Myr range shown. Water is the dominant ice constituent, and its threshold visual extinction represents the onset of the deposition of more than 1 layer of ice on the grains. At lower extinctions, the balance between desorption and accretion holds the surface coverage of the grain to less than unity. Conversion of CO to CO$_2$ is efficient at low visual extinctions, due to the higher dust temperatures, which allow CO to diffuse sufficiently rapidly to react with surface OH via reaction (2). For the high density run, the threshold visual extinction for H$_2$O is significantly lower, at around 1.5. We therefore suggest that the water (and CO$_2$) threshold visual extinctions are representative of material with a gas density of a few times $10^{3}$ cm$^{-3}$.

Figure 9 shows column densities calculated for models with various $E_{\mathrm{dif}}$:$E_{\mathrm{des}}$ ratios, adopting a gas density of $6 \times 10^{3}$ cm$^{-3}$. Panel (a) shows results for the $E_{\mathrm{dif}} = 0.35 \, E_{\mathrm{des}}$ case. Comparing with figure 8(b), we see that the slightly higher diffusion barrier moves the CO extinction threshold to lower values, while those of H$_2$O and CO$_2$ are essentially unchanged. A further increase to $E_{\mathrm{dif}} = 0.4 \, E_{\mathrm{des}}$, shown in panel (b), results in a CO$_2$ threshold in good agreement with the observed value of $6.7\pm1.6$ (Whittet et al. 2007), with the precise value dependent on the cloud age. Panels (c) and (d) show the results for $E_{\mathrm{dif}}$:$E_{\mathrm{des}}$ ratios of 0.45 and 0.5, respectively. Here, the behavior of CO and CO$_2$ changes; the lower mobility of CO in these cases renders inefficient the formation of CO$_2$ by direct diffusion and reaction with OH, leaving the main CO$_2$ formation mechanism as that in which OH forms in a binding site where CO is already present (equation 14). In these cases, the threshold of CO lies at a lower value than that of CO$_2$. Thus, only values $E_{\mathrm{dif}} \leq 0.4 \, E_{\mathrm{des}}$ are consistent with the observational data. However, higher visual extinctions may in reality have higher densities than do the edges, meaning earlier onset of CO formation and a lower threshold extinction than calculated here (i.e. closer to the observed value).

For easier comparison with observational data, linear plots of the visual extinction-dependent column densities for the cases $E_{\mathrm{dif}} = 0.3 \, E_{\mathrm{des}}$ and $E_{\mathrm{dif}} = 0.4 \, E_{\mathrm{des}}$ are shown in figure 10. While the curves for H$_2$O and CO ices are fairly linear, it may be seen that the CO$_2$ curves become less steep at higher visual extinctions, contrary to observational data (e.g. Whittet et al., 2007). This occurs at around 10 magnitudes of extinction (edge-to-edge), where appreciable CO abundances are found. This effect is explainable by the lack of dynamical considerations in the models. The visual extinction for a parcel of gas is fixed throughout its evolution, so that high visual-extinction positions never experience the low-extinction/higher-temperature conditions required to form CO$_2$ efficiently. The inclusion of dynamics in these models would likely improve the behavior at high visual extinctions, by allowing a gradual increase from low to high extinction and from high to low dust temperatures. The CO$_2$ formed at lower extinctions would be incorporated and preserved in the deepest ice mantles, allowing it to contribute to the total CO$_2$ column through the cloud later on.

While the intercepts of the linear plots for both CO$_2$ and H$_2$O are very similar for all cloud ages in the 0.5 -- 2 Myr range that is plotted in figure 10, the gradients of these curves vary with time. Whittet et al. (2007) obtain a slope for CO$_2$ of $0.252 \times 10^{17}$ cm$^{-2}$ mag$^{-1}$, using data for various lines of sight in the Taurus dark cloud. Using the $E_{\mathrm{dif}} = 0.4 \, E_{\mathrm{des}}$ model shown in figure 10(b), which has approximately the correct intercept for CO, the CO$_2$ gradient of the model over the $3 < A_{\mathrm{V}} < 8$ range best matches the observed value at a cloud age of 1.1 Myr. A similar comparison using the H$_2$O slope of $1.5 \times 10^{17}$ cm$^{-2}$ mag$^{-1}$ for the Taurus dark cloud given by Chiar et al. (1995) indicates a cloud age of 1.6 Myr. Chiar et al. also give H$_2$O column density data for the Serpens cloud. Using the same simple comparison to our model, we estimate a cloud age of 1 Myr, somewhat younger than either estimate for Taurus. Of course, without constructing a specific physical model for either cloud, such estimates are rather imprecise.

\subsection{Collapsing cores}

In order to understand the chemical behavior of the ice under collapse conditions, in which visual extinction grows and thus dust temperature falls, we employ the simple free-fall collapse model given by Brown et al. (1988) and based on the construction of Spitzer (1978). We begin collapse at a gas density $n_{\mathrm{H,0}}=3 \times 10^{3}$ cm$^{-3}$, and with a visual extinction of $A_{\mathrm{V,0}} = 2$. A peak density is chosen at which, once reached, the model remains for the duration of the run. The visual extinction evolution obeys the relationship $A_{\mathrm{V}} = A_{\mathrm{V,0}} (n_{\mathrm{H}} / n_{\mathrm{H,0}})^{2/3}$. Our adoption of a free-fall collapse model with an arbitrary maximum density is neither expected nor intended to represent the complexity of cloud-collapse dynamics, but provides a convenient treatment to understand the general behavior of the cloud chemistry under such conditions.
 
To ensure reasonable accuracy in the calculated dust temperatures at high visual extinctions, we use the dust-temperature profile calculated by Zucconi et al. (2001) (their equations 22-24 and 26) for extinctions $A_{\mathrm{V}} > 10$, and the profile derived above in section 3.2 for extinctions $A_{\mathrm{V}} < 10$. In order to allow the code easily to sample from our profile at every time step in the calculation, we fit to the $A_{\mathrm{V}}<10$ profile a third-order polynomial:

\begin{equation}
T_{\mathrm{dust}}=18.67 - 1.637 \, A_{\mathrm{V}} + 0.07518 \, A_{\mathrm{V}}^{2} -0.001492 \, A_{\mathrm{V}}^{3}\\
\end{equation}

To this profile we add a corrective of 0.316 K to ensure continuity at $A_{\mathrm{V}} = 10$. Zucconi et al. suggest their profile is valid in the $A_{\mathrm{V}}$ = 10 -- 400 range (but they do not define temperatures between 2 and 10 magnitudes of extinction, hence the use of our own approximate profile). We again use the H$_2$ and CO photodissociation treatments given by Lee et al. (1998).

Figure 11(a) shows ice mantle composition for collapse to a density of $4 \times 10^{4}$ cm$^{-3}$, which we identify with a dark cloud. This density is chosen to produce a visual extinction of approximately 10 magnitudes, which results in a dust temperature of $\sim$8.7 K. A diffusion barrier height of $0.3 \, E_{\mathrm{des}}$ is chosen. One immediately sees that CO$_2$ is formed in great abundance at early times, building up around 50 layers in a mixture mainly composed of H$_2$O and CO$_2$; CO abundance is minimal in this regime. CH$_4$ also reaches its peak fractional composition in these layers, as does ammonia. CO$_2$ reaches its peak at around $7 \times 10^5$ yr, at which the dust temperature is 12 K. After this, there is a rapid switch-over in behavior, and CO dominates over CO$_2$. We identify the CO$_2$ formed in layers 0 -- 50 with the observed polar CO$_2$ component. At temperatures of less than 12 K, CO diffusion becomes unable to compete with hydrogenation of OH, and so the alternative mechanism of OH formation in the presence of CO becomes the most important for CO$_2$ production.

Figure 11(b) shows the gas-phase abundances for the same model, also mapped to the number of ice layers. Gas-phase atomic carbon is seen to fall off at early times, as it is converted to CO at relatively low gas densities. Initially high gas-phase carbon abundances produce the early CH$_4$ ice peak. This declines with increasing rapidity as the gas density rises, accelerating the conversion of gas-phase C to CO. This coincides with the sharp fall in CO$_2$, also caused by rising density and visual extinction, and thus dust temperature. These two species should be well correlated in ice observations. This behavior differs from the static dark-cloud models, which forced gas-phase CO to form concurrently with the ice mantles, producing high methane content in the ice. Under collapse conditions, much atomic carbon is converted to CO before significant ice build-up occurs. Even longer periods at low density/$A_{\mathrm{V}}$ prior to collapse would allow more carbon to be locked into CO, resulting in yet lower methane ice abundances.

It is also seen that methanol, formed on the grains, maintains an appropriate gas-phase abundance, in the region where CO dominates CO$_2$ in the ice mantles. Therefore, in regions of visual extinction less than the CO threshold, no appreciable gas-phase methanol is to be expected.

Figure 11(c) shows the cumulative fractional composition of the ice summed through the entire ice mantle (up to a thickness of the value shown on the abscissa), with respect to water. These values represent only the ice composition within a single parcel of cloud gas, and do not take into account any integration through the outer parts of the cloud. For an ice thickness of around 50 layers, cumulative CO$_2$ fraction is at its peak, comprising approximately 80 \% of the water abundance. This value falls for greater ice thicknesses, eventually falling below the total CO fraction. The last column of Table 3 shows the ``best-match'' results for this model as compared to the dust-grain ice composition for Elias 16 presented by Gibb et al. (2000). We roughly determine the ``best-match'' as the time at which both CO and CO$_2$ comprise 30 \% of the ice mantle, which occurs for ice thickness of $\sim$220 layers. The new values are a much better match, even for CH$_4$, which was problematic for the static dark-cloud models. The abundance of NH$_3$ is nevertheless rather high for dark-cloud observational values, although it does not exceed levels observed toward star-forming regions (20 -- 30 \%, Gibb et al., 2000), after its initial peak. The introduction of collapse from a lower visual extinction also allows a much better reproduction of observed CO$_2$ values. The values determined by the model are of course somewhat dependent on the choice and complexity of the dynamical model.

Figure 12 shows the results for collapse to a density of $10^7$ cm$^{-3}$, which is achieved after around $9.5 \times 10^5$ yr, and by which time extreme depletion of gas-phase species has occured. The general composition and behavior of the ice is not greatly different from the lower density run; however, methanol formation falls dramatically after its peak at around 50 layers. This is caused by the much lower dust temperature ($\sim5$ K), which allows H$_2$ to maintain a dominant level of ice-surface coverage. The high deposition rate of H$_2$ onto the ice reduces the period of time available for atomic hydrogen to hydrogenate CO and H$_2$CO to methanol before a new layer of ice is deposited on top. Thus, the formation of significant quantities of methanol in very dense cores, with such low temperatures, is unlikely. However, it is not clear whether dust temperatures this low could be achieved before star formation were to warm the dust grains, and the free-fall collapse model produces only a simplistic treatment of physical conditions.

\section{Discussion}

These models show that CO$_2$ may be formed on dust-grain surfaces by the CO + OH $\rightarrow$ CO$_2$ + H reaction alone. The efficiency of this barrier-mediated reaction is determined by the competition between reaction and the loss of the reactants by their diffusion out of the surface binding site. Using our standard surface-binding characteristics, we find the efficiency to be close to unity. Under such a regime, the specific mechanism and rate at which CO and OH meet on the surface are the determining factors for CO to CO$_2$ conversion. At sufficiently high temperatures ($\sim$12 K), CO is mobile enough to reach and react with OH radicals before either H or H$_2$, producing a high CO-to-CO$_2$ conversion efficiency and allowing CO$_2$ to dominate CO in the ice mantles. At lower temperatures, CO$_2$ is still formed in significant quantities; however, in this case the OH is formed atop a CO molecule by the addition of O and H, and CO abundance always dominates that of CO$_2$. Typical gas-grain codes whose reaction rates are determined by diffusive processes do not take such a mechanism into account. Importantly, the only diffusion required for this process is that of atomic hydrogen; oxygen may be hydrogenated wherever it lands on the grain surface. While the release of chemical energy associated with the formation of OH is likely to render the CO + OH reaction more rapid (perhaps increasing the probability of the dissociation of the excited intermediate HOCO* into CO$_2$ and H), in this model it is simply the act of bringing CO and OH together that allows the mechanism to be so efficient.

In contrast to reaction (2), reaction (3) is rendered ineffective by the calculated quantum tunneling rates of Goumans \& Andersson (2010), while reaction (1) is never competitive, due to the relatively low mobilities and abundances of each reactant (O and HCO).

It is of note that time-dependent changes in the gas-phase chemistry of static dark-cloud models do not appear able to induce a switch-over from CO$_2$ to CO dominance in the ice mantles; it would seem that a change in physical conditions is required for the CO:CO$_2$ ratio in the ice mantles to be drastically altered. While the gas density does not appear to have a direct effect on the CO:CO$_2$ ratio, adjustments to the dust temperature are capable of inducing a switch-over, by changing the dominant mechanism by which CO and OH meet on the grains, as mentioned above. However, at temperatures in the 8 -- 12 K range we find that the general behavior of the ice chemistry is remarkably robust; previous models (e.g. Ruffle \& Herbst 2001) have often shown strong variation with temperature, due to the use of the simple deterministic ``rate-equation'' approach to solve the surface chemical network.

Our simple treatment of the dependence of dust temperature on visual extinction allows the observational threshold extinctions of H$_2$O and CO$_2$ to be well reproduced, assuming a gas density of a few $\times 10^{3}$ cm$^{-3}$. Higher densities tend to lower the threshold extinction for these species. The threshold extinction for CO can also be reproduced by the models, but the precise value is dependent on the ratio of diffusion rates to binding energies on the grain/ice surfaces, and is also more sensitive to the assumed age of the cloud than those for the other two molecules. The diffusion barriers determine the temperature at which surface CO becomes sufficiently mobile to induce a switch over in CO:CO$_2$ dominance.  However, using the adopted dust temperature profile, the barriers must be less than $0.4 \, E_{des}$ to be able to produce the switch-over at all. Somewhat lower values than this cannot be ruled out, as the use of a more realistic physical model -- either through collapse or through a non-uniform density profile -- may also cause the modeled CO:CO$_2$ boundary to shift. 

The observational extinction threshold for H$_2$O corresponds to the point where the photodesorption of water (and other surface species) back into the gas phase is not sufficient to hold the ice to less than one layer of surface coverage. As the rate of ice formation increases further, the total photodesorption rate is held steady (in absolute terms), since material can only desorb from the surface and not from deeper layers. This effect will remain essentially the same even if photodesorption occurs from the first few surface layers. 

CO$_2$ formation occurs as soon as gas-phase CO is present. In our low-density collapse models, a few layers of water ice are formed before gas-phase CO becomes abundant. However, a collapse model that began at a visual extinction lower than 2 magnitudes would show CO formation before any significant ice mantles were formed. The slightly higher observed extinction threshold for CO$_2$ than for H$_2$O probably derives from the slightly lower abundance of CO$_2$ in the ices.

In the case of CO, the extinction threshold is tied to a temperature threshold of around 12 K, at which point CO mobility is insufficient for reaction (2) to occur via direct diffusion. As a collapsing cloud or core reaches its threshold visual extinction, the dust becomes cool enough that the less efficient mechanism takes over.

The efficient formation of CO$_2$ by the addition of CO and OH via direct surface diffusion is dependent on the reactants having sufficient time in the binding site to react before one of them diffuses away. If the barrier to the first stage of this reaction, corresponding to CO + OH $\rightarrow$ HOCO*, is low enough enough then formation of CO$_2$ {\em may} occur. There is some debate as to the precise value of this barrier, however it is typically estimated to be low enough that our assumption of high efficiency for reaction (2) is very plausible. As discussed by Goumans et al. (2008), the rapidity with which excited HOCO* loses energy to the grain surface may be important to the mechanism by which CO$_2$ is ultimately formed. But, the abundance of gas-phase H combined with its rapid surface diffusion make quite probable the alternative process of H-addition to de-excited HOCO that those authors suggest. A more detailed chemical model would be valuable in confirming this expectation. Goumans et al. (2008) also suggested that HCOOH would be another significant product branch for HOCO hydrogenation. No alternative product branches are included in our network. However, Oba et al. (2010) detected no such signatures, perhaps indicating that the loss of energy to the surface is not sufficiently fast to affect the direct dissociation of HOCO to H + CO$_2$. All of these considerations are important in the case of direct diffusion of CO to OH, but the mechanism by which O and H combine on top of a CO molecule would involve a highly excited OH radical which could overcome both any initial barrier to HOCO* formation and the internal barrier to the ejection of an H atom to produce CO$_2$.

The formation of CO$_2$ at low visual extinctions, as demonstrated by the models, requires a certain maximum diffusion-barrier height of $E_{\mathrm{dif}} \simeq 0.4 \, E_{\mathrm{des}}$ for CO, although here we change the generic $E_{\mathrm{dif}}$:$E_{\mathrm{des}}$ ratio to avoid excessive parameter space. It is possible that higher values could produce the same effect, under higher dust-temperature conditions, although our simple extinction-dependent temperature profile agrees reasonably well with the much more precise approach of Zucconi et al. (2001) whose data we use to make our approximation. However, those authors do not give well-defined temperatures in the $A_{\mathrm{V}}$=2 -- 10 range. A departure from the profile we assume could thus affect the threshold extinction for CO produced by the models.

CH$_4$ is generally less abundant than NH$_3$ in the ices due to the gas-phase conversion of atomic carbon to CO, where atomic nitrogen does not have a similarly rapid removal process. Water production is not affected in the same way as CH$_4$, due to the larger gas-phase abundance of atomic oxygen. In the case of a collapsing core, CH$_4$ and CO$_2$ are correlated in the ices (a result also suggested by the observational data of {\"O}berg et al., 2008), due to the formation of CH$_4$ at early times, and the formation of CO$_2$ at the high temperatures and low visual extinctions of the pre-collapse stage.

Observations indicate that a large proportion of interstellar CO$_2$ ice is embedded in a polar mixture (e.g. Gibb et al. 2000, Bergin et al. 2005), i.e. a water-rich ice, and this component of the ice mantles may be easily explained by the efficient conversion of CO to CO$_2$ at low visual extinctions and elevated temperatures. In dark clouds, this component would thus be formed at dynamically early times, before the gas density and visual extinction had reached more typical dark-cloud values. Furthermore, observations toward the Taurus field star Elias 16 by Bergin et al. (2005) found that, while the majority of detected CO$_2$ ice is in a polar mixture, around 15 \% resides in a non-polar CO:CO$_2$ ice dominated by CO; indeed, their best match to this ice composition was a 100:26 mixture. The static dark-cloud models we present here typically produce a CO:CO$_2$ ratio of approximately 4:1 to 2:1, dependent on the physical parameters of the model -- this ratio is very close indeed to the (non-polar) observational result, and is fairly robust through many layers of ice. However, at all times in those model runs, water is the most abundant ice species. Indeed, there is no obvious means by which one can engineer the models to produce both a low water content and an appropriate CO:CO$_2$ ratio that endures for a sufficent time. Under the dominant CO$_2$ formation mechanism of the dark-cloud models, in which OH is formed in the presence of CO, a significant amount of water is always formed, as the price of the inefficiency of CO to CO$_2$ conversion that produces the CO:CO$_2$ ratio of $\sim$2 -- 4. It is possible that mobile atomic oxygen could become more strongly bound to CO, such that they remained together until hydrogenation resulted in the formation of CO$_2$, as suggested by Goumans \& Andersson (2010). In such a case, the CO:CO$_2$ ratio could be explained by the relative gas-phase abundances of CO and atomic oxygen, resulting from depletion, while water-ice abundances could remain low. However, such a depressed gas-phase oxygen abundance is likely to be a transient feature of the chemistry prior to even greater depletion, as shown by figure 2(b), while the severely depressed gas-phase atomic hydrogen abundances required to allow oxygen sufficient time to diffuse to the nearest CO molecule on the grain surface without being hydrogenated are unlikely to pertain in the moderate density conditions of dark clouds.

We therefore suggest that the observed apolar characteristics of CO$_2$ (and indeed of CO) may not be indicative simply of water-poor ice; rather, they would indicate that the CO$_2$ is formed from a highly-segregated CO:H$_2$O mixture. In such a scenario, CO molecules would form islands within a predominantly water-ice composition. The formation of OH on these CO-rich islands would lead to CO$_2$ formation on the island, while the immobility of any OH formed on a water surface would ultimately lead to its hydrogenation, producing more water. This structure would be incorporated into the deeper ice layers as new surface layers were formed, producing ``lumps'' of CO within the H$_2$O-dominated mantle. {\"O}berg et al. (2009c) considered the segregation of ice mixtures in protostellar envelopes, caused by the preferentially higher binding of molecules to each other than to different species. They suggested a barrier to CO:H$_2$O segregation of $300 \pm 100$ K, although the relative importance of surface or bulk diffusion or swapping processes was not well defined. In our dark-cloud regime, the CO and H$_2$O molecules would segregate {\em as the ice mantle formed} by deposition of material from the gas phase. While CO is not sufficiently mobile at around 10 K for reaction (2) to occur through direct diffusion, it {\em is} mobile enough to find other surface CO molecules (which are relatively abundant, at around 10 \% surface coverage), before a new layer of ice is formed and the structure is more firmly locked in. Even assuming a dust temperature of 9 K, with surface diffusion barriers as high as $0.4 \, E_{des}$ (giving $E_{dif}($CO$)=460$ K), grain-surface CO molecules could (under dark cloud conditions) manage around 10 thermal hops before a new layer of ice formed -- just sufficient to segregate effectively. As atomic oxygen has to be on top of a CO molecule for CO$_2$ to form, all CO$_2$ formed at low temperatures is likely to be adjacent to many CO molecules, producing the observed spectral characteristics. Under the high-temperature regime, CO moves to OH, giving an essentially random arrangement of CO$_2$, which is thus well mixed with H$_2$O. Even at temperatures as high as 20 K, the surface mobility of CO$_2$ is not sufficient to re-arrange itself before a new ice layer has been deposited. These suggestions are speculative; however, the requirement that the accretion rates be small compared to rates of surface diffusion could make them difficult to test directly in the laboratory.

The above scenario is also sufficient to explain the generally low abundances of polar CO ice observed in dark clouds (e.g. Gibb et al. 2000); large islands of CO surrounded by water ice would produce a predominantly apolar signature.

The collapse models show that, under the simple dynamical model we employ, the first 50 or so layers of ice are extremely rich in CO$_2$, while the CO$_2$ formed following the abrupt switch-over to CO dominance is around 5 -- 8 times less abundant. On this basis, the 15 \% polar CO$_2$ found by Bergin et al. (2005) toward Elias 16 would suggest a cloud age of around 8 -- $9 \times 10^{5}$ yr, using the $n_{\mathrm{H,max}}=4 \times 10^{4}$ cm$^{-3}$ model. The model CO$_2$ profiles are also in general agreement with the collated ice-observation data presented by Gibb et al. (2000), in which polar CO$_2$ is dominant over apolar (although comparison with the other detected ice species suggest an age a little over 1 Myr). The collapse models would suggest that most of the CO$_2$ observed toward high-extinction objects is formed during an early period of low extinction. A more extended low-density period would also improve the comparison with CH$_4$ ice abundances. The free-fall collapse models that we use are naturally very approximate; more detailed dynamical models would likely reproduce specific features of the ice composition more accurately.

A further consequence of the CO$_2$-formation scenario we suggest above is that different grain-size populations could exhibit different CO:CO$_2$ ratios due to their differing temperatures, which should scale as $a^{-1/6}$ (see section 3.2). Indeed, due to the sharp switch-over between CO$_2$ regimes at around 12 K, one might expect a bi-modal distribution of polar/apolar characteristics with grain-size, assuming static dark cloud conditions. However, the changing dust temperatures engendered by the collapse process would complicate this picture, as each grain population would acquire both polar and apolar ices during their evolution. It is also not clear how the chemistry would be affected by differing grain sizes, as not only the dust temperatures but their accretion cross-sections and surface diffusion rates would also be altered. The threshold temperature of $\sim12$ K could be affected by such changes. These questions can only be addressed by the use of more comprehensive gas-grain chemical models.

The three-phase model that we employ here allows the ice composition formed at any stage of cloud evolution to be preserved through later stages. As a result we find that, contrary to the findings of Garrod et al. (2006), most carbon ends up as CO, CO$_2$, CH$_4$ or CH$_3$OH, not as large hydrocarbons. Because only the surface ice layer is active either chemically or physically, the dissociation or re-injection into the gas phase of ice material is strongly limited, which could otherwise cause hydrocarbon dominance in the ices.

The formation of significant quantities of methanol ice requires that the deposition of new ice layers be sufficiently slow for atomic hydrogen to have time to react with CO and H$_2$CO before they are incorporated into the deeper ice mantles. In the dark-cloud case, this occurs when atomic oxygen begins to deplete substantially from the gas phase, reducing the rate of water-ice production, and removing a sink for surface atomic hydrogen. This also tends to occur at the same time as CO becomes the dominant gas-phase species other than H$_2$, as a result of the steady conversion of C and O into CO. In the case of a dense collapsing core, we find that very low dust temperatures and/or high density conditions lead to the rapid deposition of new ice layers; molecular hydrogen deposition becomes especially important in very cold, dense conditions. This leaves CO very abundant in the outer ice-mantle layers under collapse conditions, with high-density cores achieving minimal conversion to methanol. Thus, the model suggests either that dense and/or collapsing cores that are rich in methanol may have spent a significant period at intermediate densities and temperatures (at which hydrogenation is more efficient), or that the process of star formation has elevated the dust temperature, such that H$_2$ coverage is insignificant. Such conditions could explain the particularly high methanol-ice fractions detected in low-mass star-forming cores (Boogert et al. 2008). However, the models presented here say nothing about the chemistry in such sources where significantly higher temperatures or strong UV radiation fields obtain as a result of the star-formation process. It should also be noted that the activation barriers and tunneling rates for the hydrogenation of CO and H$_2$CO are somewhat uncertain.

It is clear that, in order to treat ice chemistry accurately, models are required that can distinguish the surface of the ice from the underlying mantle. This is especially important for low temperature regimes, where mobility within the mantles is likely to be very low, and the penetration of surface reactants into the mantle is minimal. It is also necessary to use a physically accurate treatment of the efficiency of barrier-mediated reactions, and, in the case of CO$_2$ production, to include an indirect formation mechanism to account for the formation of OH in the same binding site as a CO molecule. Much more accurate determination of various quantum-tunneling rates for barrier-mediated reactions would improve model accuracy. Furthermore, models such as this one that can simulate both the grain-surface and gas-phase chemistry are necessary to fully understand the evolution of dust-grain ices in interstellar environments; the accretion rates of gas-phase species strongly affect the grain-mantle chemical composition. For an accurate treatment of CO$_2$ ice formation in dark clouds, the inclusion in the model of some form of collapse process also appears to be necessary.

\section{Conclusions}

We list the main conclusions of this modeling study below:

\begin{enumerate}

\item The reaction CO + OH $\rightarrow$ CO$_2$ + H is shown by our models to be efficient enough to account for the vast majority of CO$_2$ formation. If OH is formed on top of a CO surface, the only surface diffusion required to produce CO$_2$ is that of atomic hydrogen.

\item At low temperatures, the grain-surface formation of OH on top of a CO molecule allows CO$_2$ to be formed in a ratio CO:CO$_2$ $\simeq$ 2 -- 4, under dark cloud conditions. These ratios are robust over a wide range of temperatures below $\sim$12 K. We identify this component with observed apolar CO and CO$_2$ signatures in dark clouds.

\item At higher temperatures ($> 12$ K), CO is sufficiently mobile to compete effectively for reaction with OH, producing ice mantles rich in CO$_2$ and very poor in CO. We identify this component with observed polar CO$_2$ signatures in dark clouds.

\item Visual extinction-dependent chemical models appear capable of reproducing the observed extinction-threshold behavior of H$_2$O, CO$_2$, and CO. 

\item The decline in dust temperature caused by the extinction of the interstellar radiation field at higher visual-extinctions causes the change in CO and CO$_2$ chemical behavior on the grains. A sharp switch-over occurs in the CO:CO$_2$ ratio at a temperature of around 12 K. The visual-extinction threshold of CO is thus tied to this temperature threshold.

\item The water-ice extinction threshold is well reproduced with a model gas density of a few $\times 10^3$ cm$^{-3}$. This threshold is associated with the rate of photodesorption from the grain surface. CO$_2$ ice formation occurs in tandem with that of H$_2$O, so long as CO is present in the gas phase.

\item The observed polar CO$_2$ is associated with early-time formation in tandem with water ice. We suggest that the apolar component is associated with CO$_2$ that exists within a highly segregated CO:H$_2$O ice. This segregation would occur during the formation of each ice layer, producing CO-rich lumps within the H$_2$O. Formation of OH upon the CO surface would produce CO$_2$ that is trapped on that surface and would be incorporated into the CO-rich ice. A small polar CO signature could be produced at the interface of the CO with the H$_2$O.

\item Variations in dust temperature across the canonical grain-size distribution may be important in determining polar/apolar ice ratios, due to the sharp temperature dependence of CO/CO$_2$-formation rates on the grains.

\item In the dark-cloud models, large methanol-ice abundances with respect to water are indicative of the depletion of gas-phase oxygen atoms at late times. These large methanol abundances reside in the outer layers of the ice mantle.

\item Under high-density and high-extinction conditions, extremely cold temperatures lead to high surface coverage of molecular hydrogen, resulting in less methanol production due to the shorter times available to hydrogenate before new layers are deposited. Thus, ice mantles in cold, dense, collapsing cores are CO-rich in all but the deepest layers. The efficiency of methanol formation at high densities is extermely sensitive to timescale and dust temperature.

\item Gas-phase methanol is solely produced (on the grains) during the later, apolar CO/CO$_2$ phase, which occurs after the CO extinction threshold has been exceeded. The gas-phase abundance is highly stable over this period.\\

\item Methane (CH$_4$) and CO$_2$ in the ice mantles appear to be well correlated, due to the coincidence of high gas-phase atomic carbon abundances (in the case of CH$_4$) and elevated dust temperatures (in the case of CO$_2$). Methane and CO$_2$ ices are thus mainly associated with the deeper ice layers in the mantle. The very low observed CH$_4$ ice abundances are indicative of long periods of conversion of C to CO in the gas phase, at visual extinctions in the range 1 -- 2 (from cloud edge to center), prior to any collapse. 

\item More detailed gas-grain models that deal with the short-lived intermediates of CO$_2$ production would allow a closer investigation of the relevant reaction processes and would improve model accuracy. The incorporation of a grain-size distribution into the models will also be important to the explanation of observational results.

\end{enumerate}

\

\acknowledgements
We thank T. Goumans for many helpful discussions on the tunneling rates of barrier-mediated reactions. We thank K. {\"O}berg for helpful discussions on ice spectra. We also thank the referee, S. Charnley, for helpful suggestions on the manuscript. This work was partially funded by a grant from the NASA Astrophysics Theory Program. TP thanks the National Science Foundation for a studentship under the Research Experiences for Undergraduates (REU) program, which was undertaken at Cornell University. 


\begin{figure}
\includegraphics[width=0.40\textwidth]{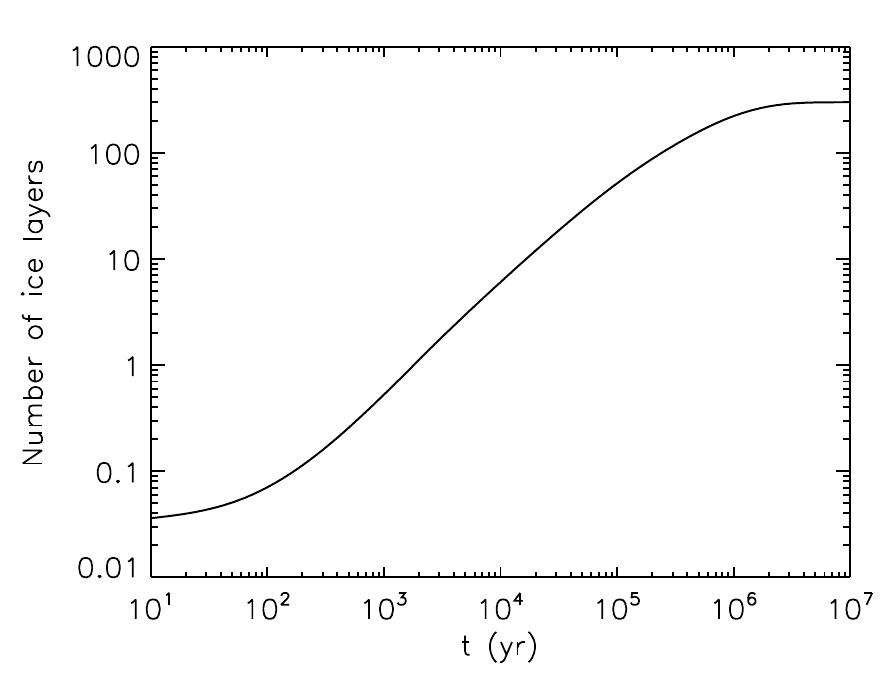}
\caption{Growth of ice mantle over time, for static dark-cloud model at $T_{\mathrm{gas}}=T_{\mathrm{dust}}=10$ K.}
\end{figure}

\begin{figure}
\includegraphics[width=0.475\textwidth]{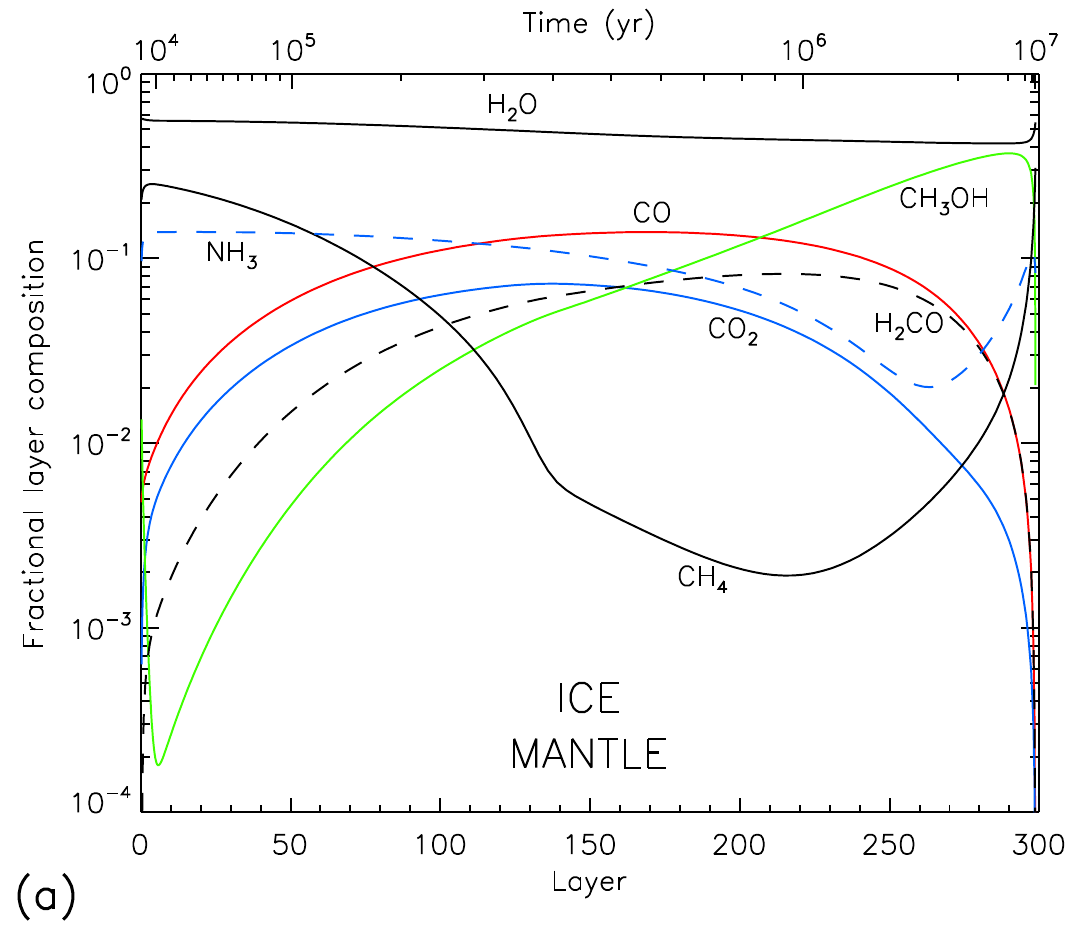}
\includegraphics[width=0.475\textwidth]{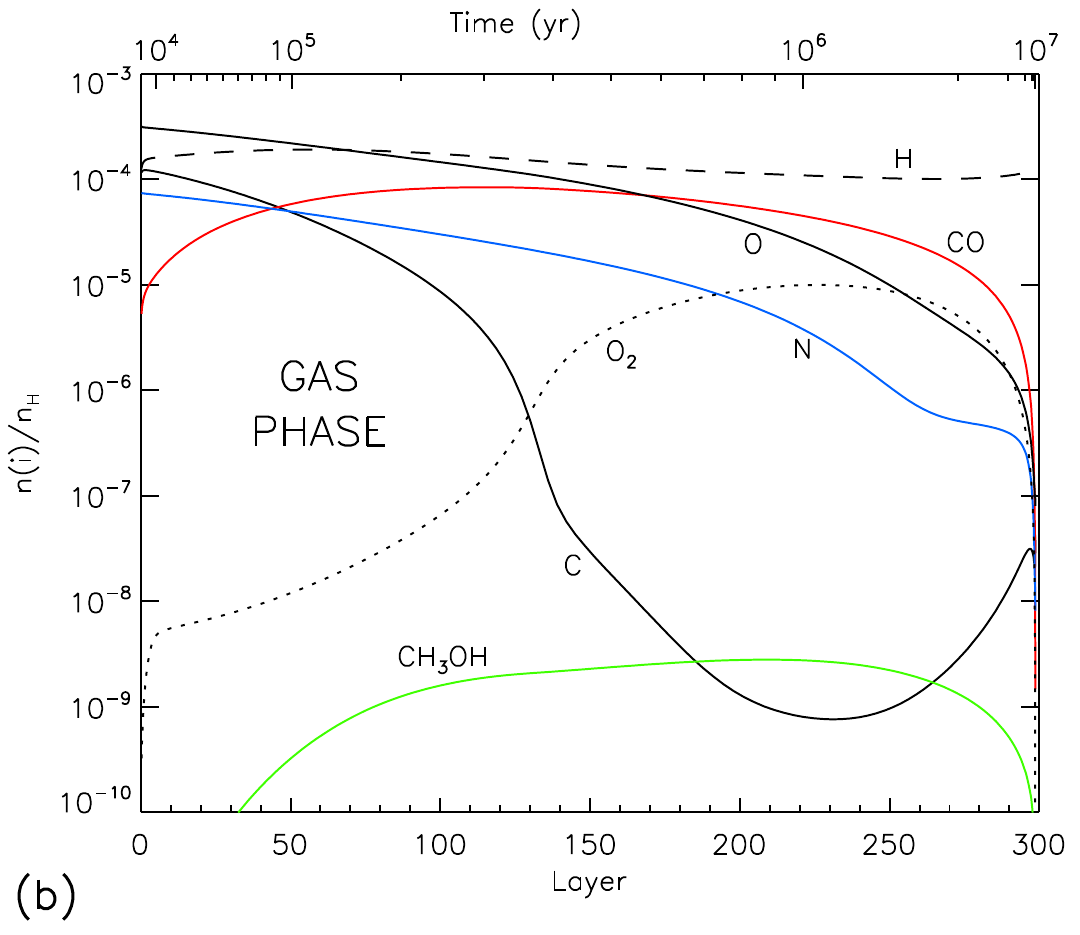}
\caption{Basic dark-cloud model at 10 K; {\bf a)} Fractional ice composition; {\bf b)} Gas-phase abundances with respect to total hydrogen.}
\end{figure}

\begin{figure}
\includegraphics[width=0.475\textwidth]{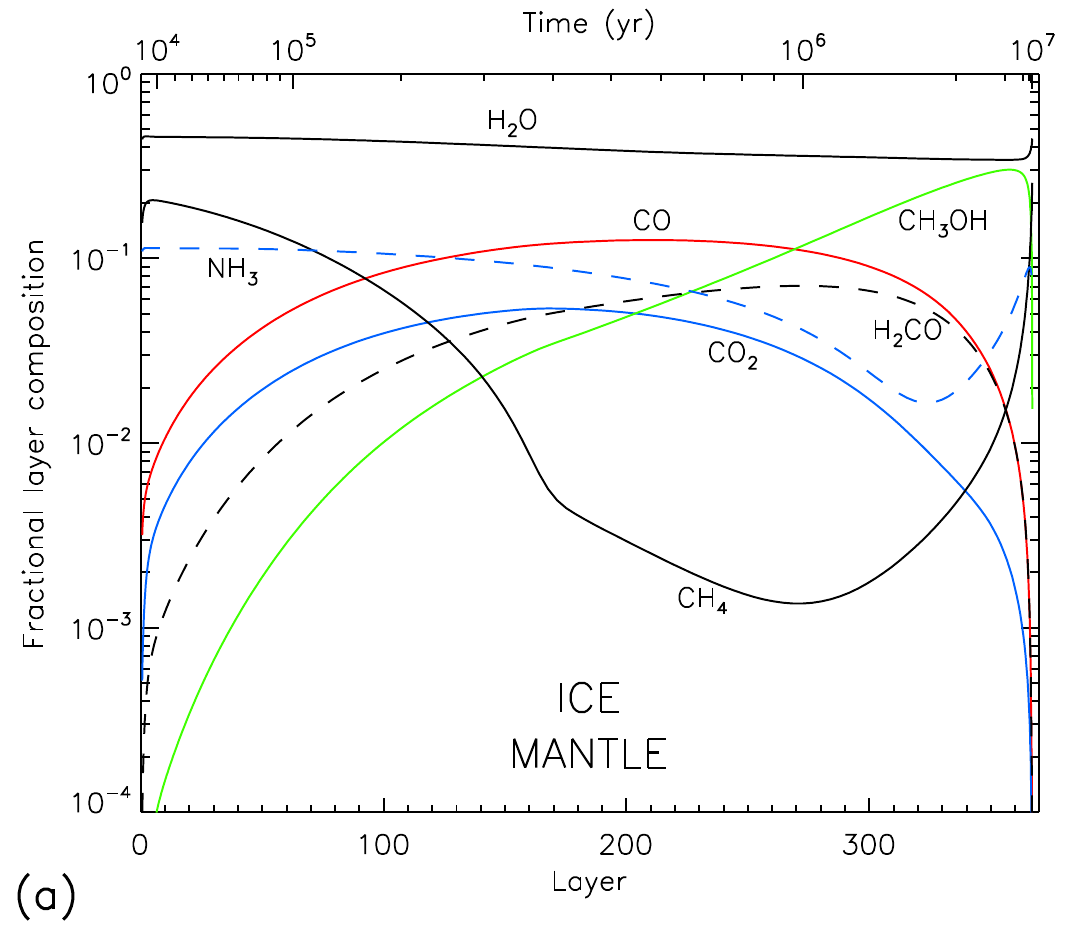}
\includegraphics[width=0.475\textwidth]{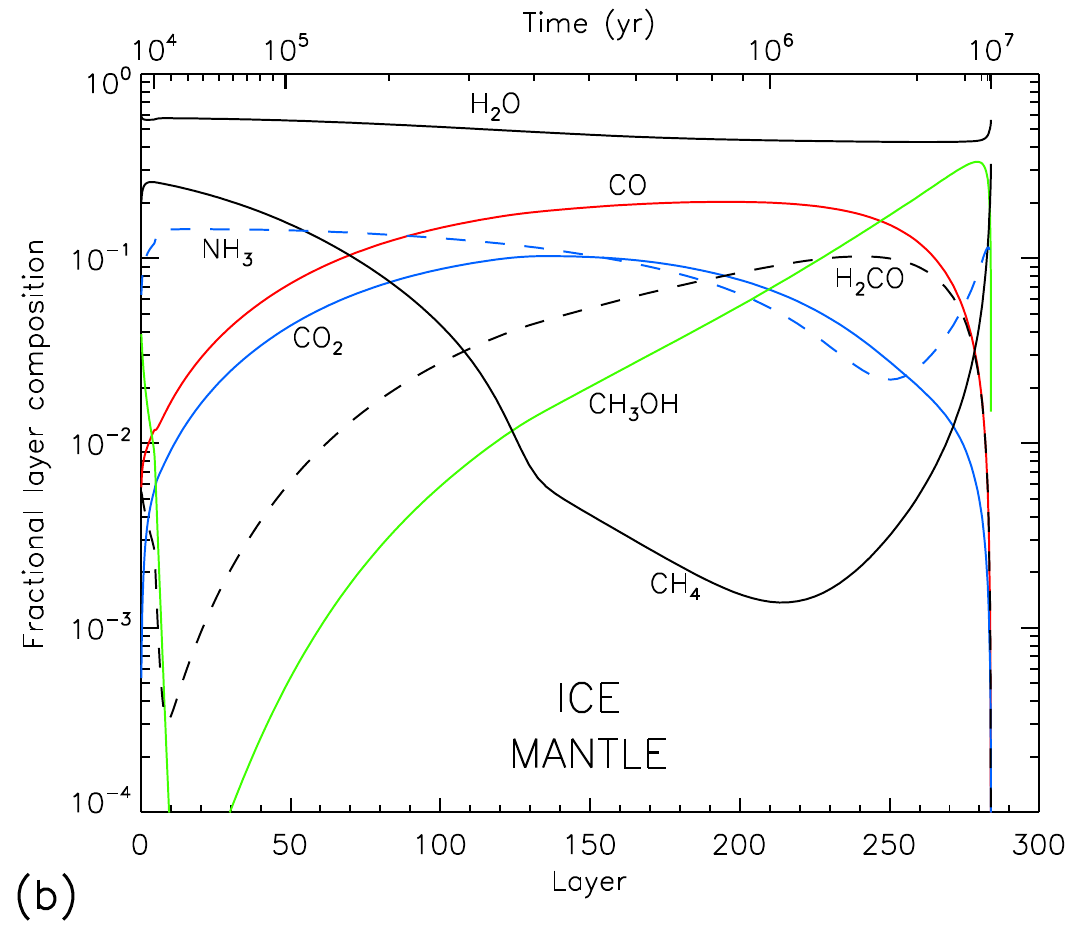}
\caption{Ice-mantle composition by layer, with dust-grain temperatures: {\bf a)} 8 K; {\bf b)} 12 K. Gas temperature is held at 10 K for all runs.}
\end{figure}

\begin{figure}
\includegraphics[width=0.475\textwidth]{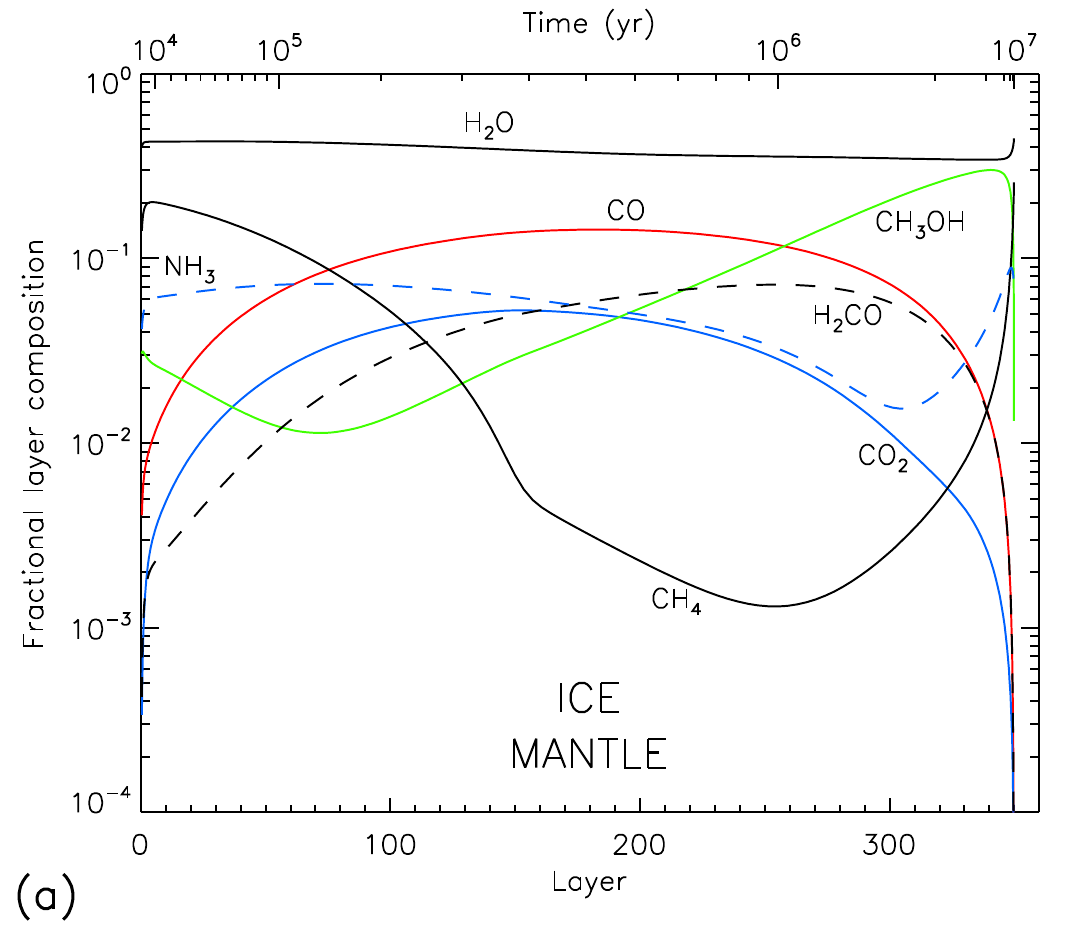}
\includegraphics[width=0.475\textwidth]{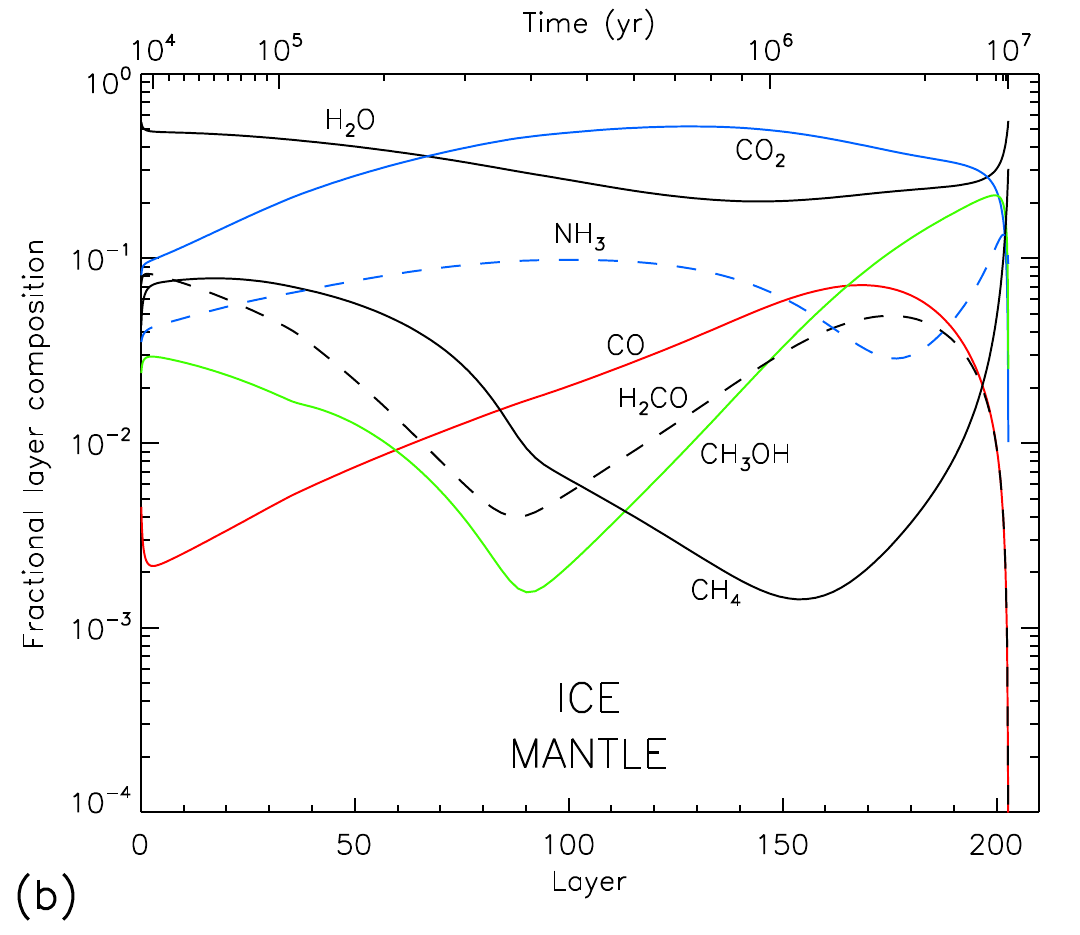}
\caption{Ice-mantle composition by layer, assuming $E_{\mathrm{dif}}= 0.3 \, E_{\mathrm{des}}$, with dust-grain temperatures: {\bf a)} 8 K; {\bf b)} 12 K. Gas temperature is held at 10 K for all runs.}
\end{figure}

\begin{figure}
\includegraphics[width=0.475\textwidth]{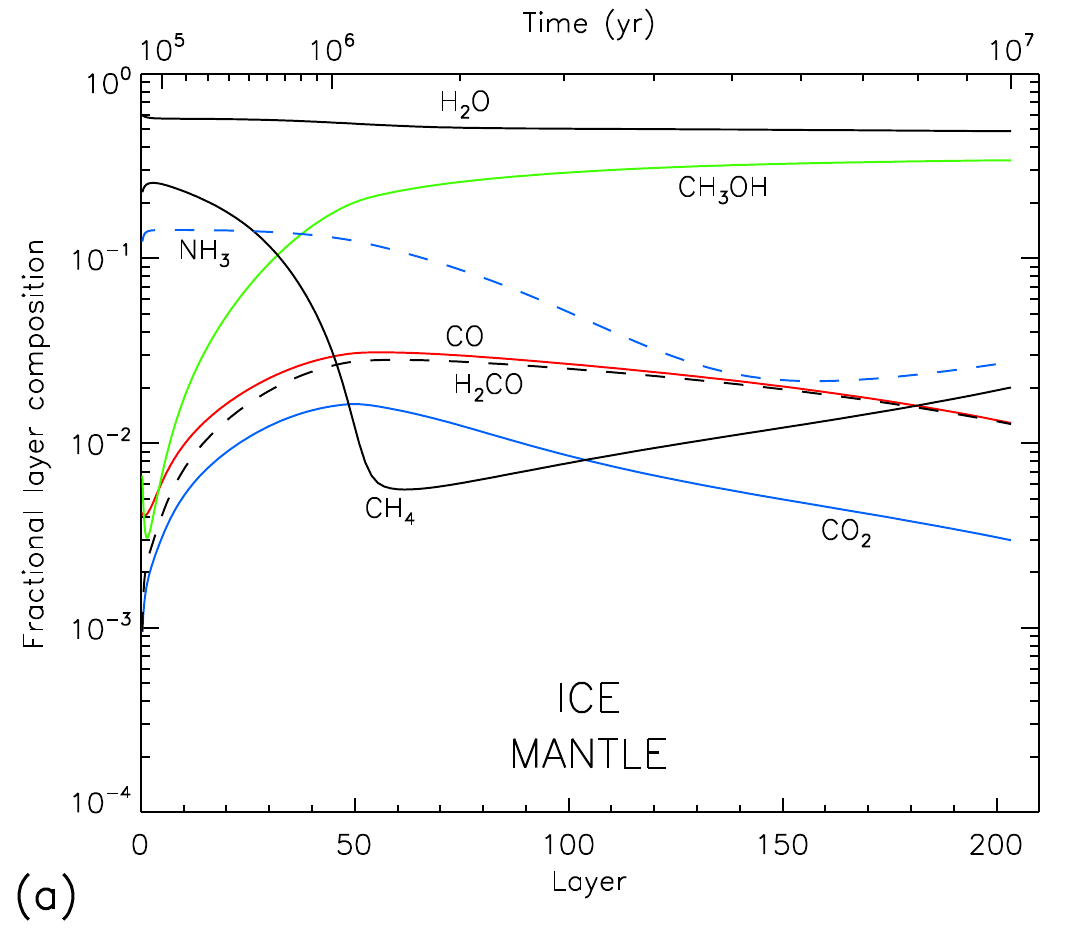}
\includegraphics[width=0.475\textwidth]{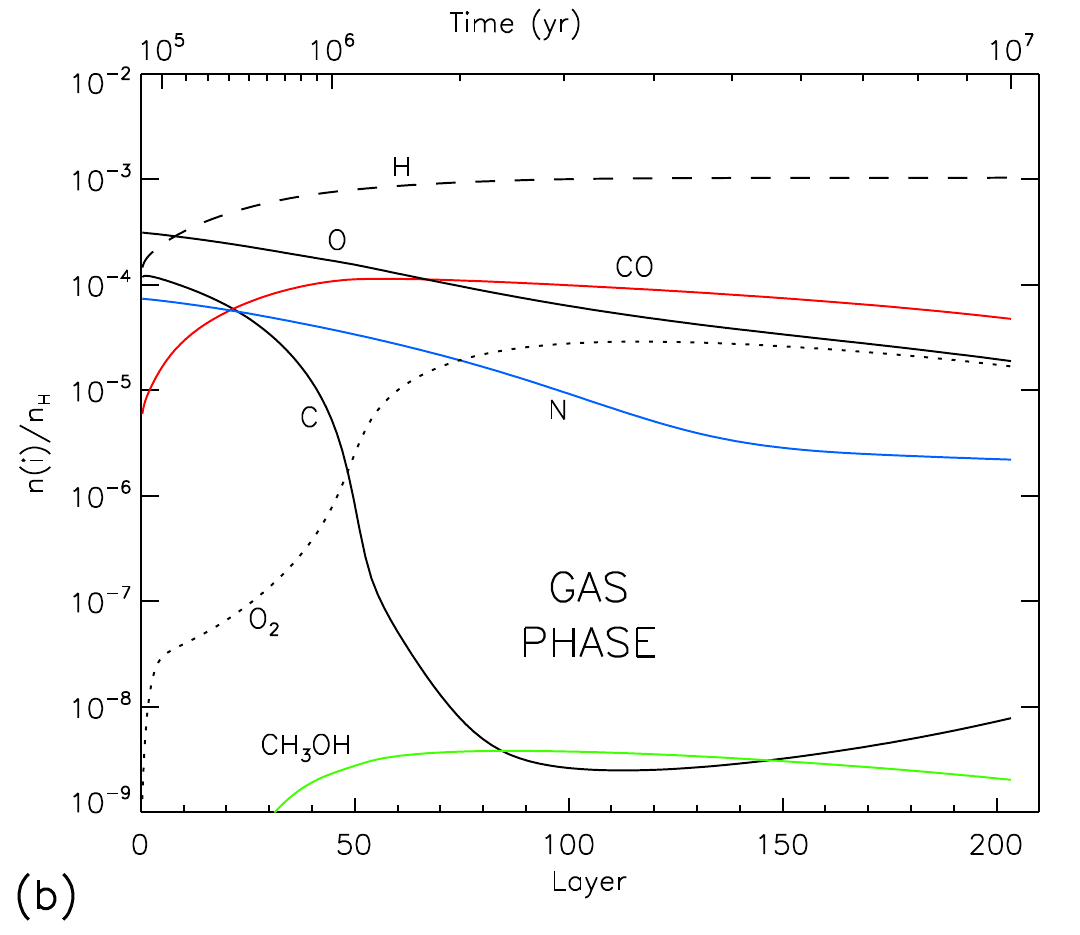}
\caption{Ice-mantle composition by layer, assuming $n_{\mathrm{H}}=2 \times 10^{3}$ cm$^{-3}$, panels as per figure 2.}
\end{figure}

\begin{figure}
\includegraphics[width=0.475\textwidth]{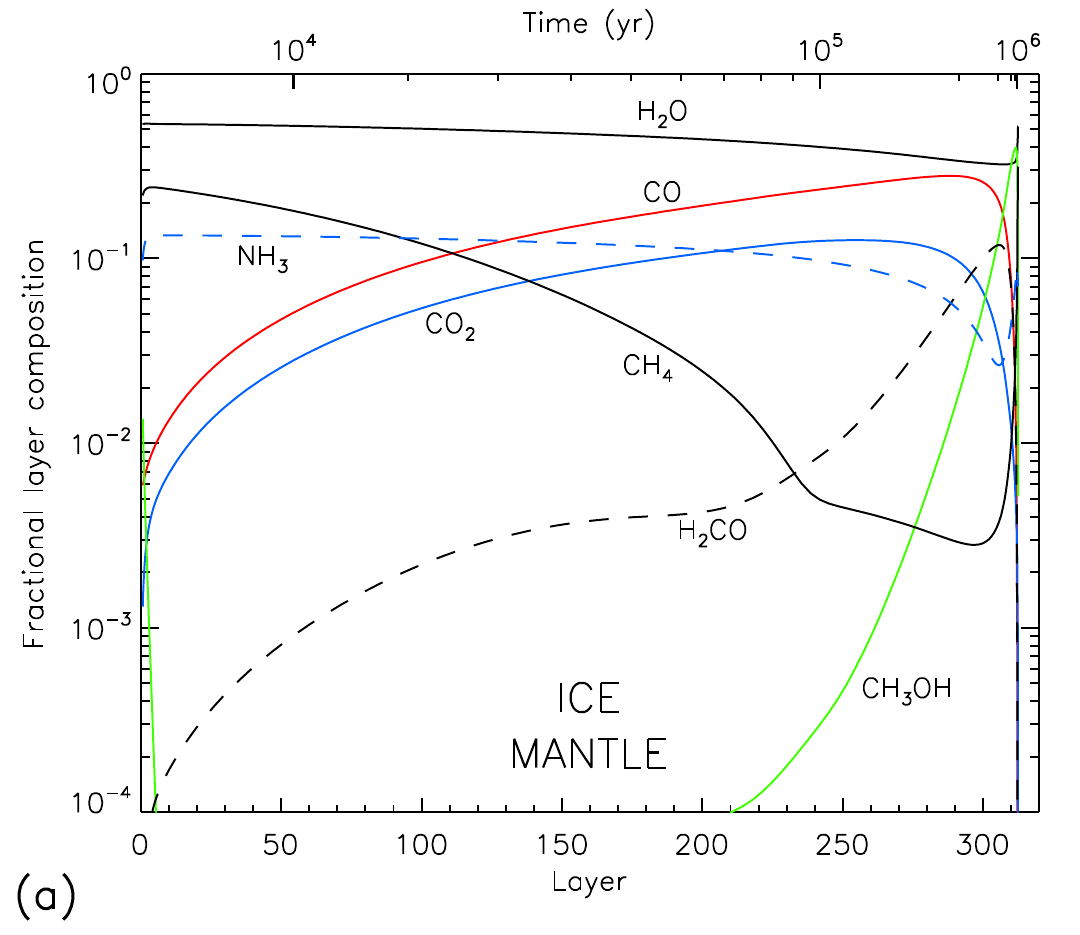}
\includegraphics[width=0.475\textwidth]{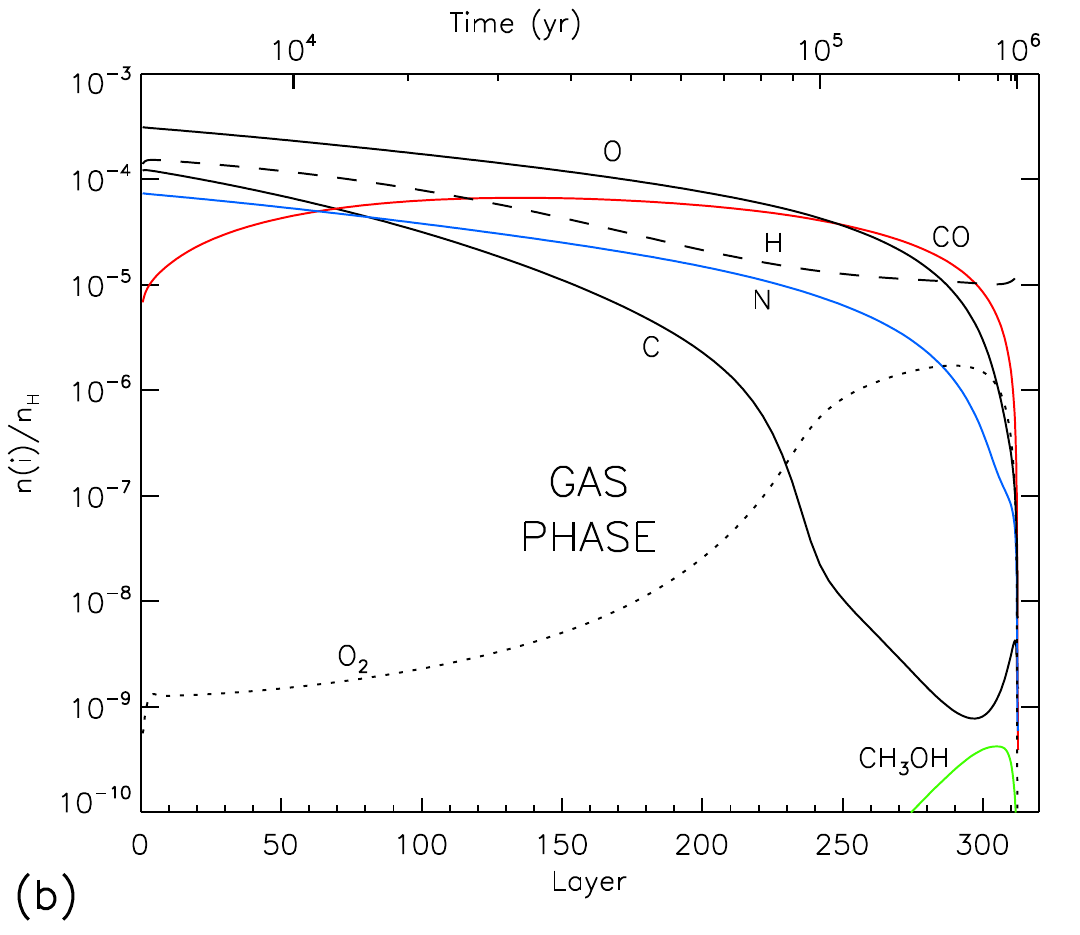}
\caption{Ice-mantle composition by layer, assuming $n_{\mathrm{H}}=2 \times 10^{5}$ cm$^{-3}$, panels as per figure 2.}
\end{figure}

\begin{figure}
\includegraphics[width=0.40\textwidth]{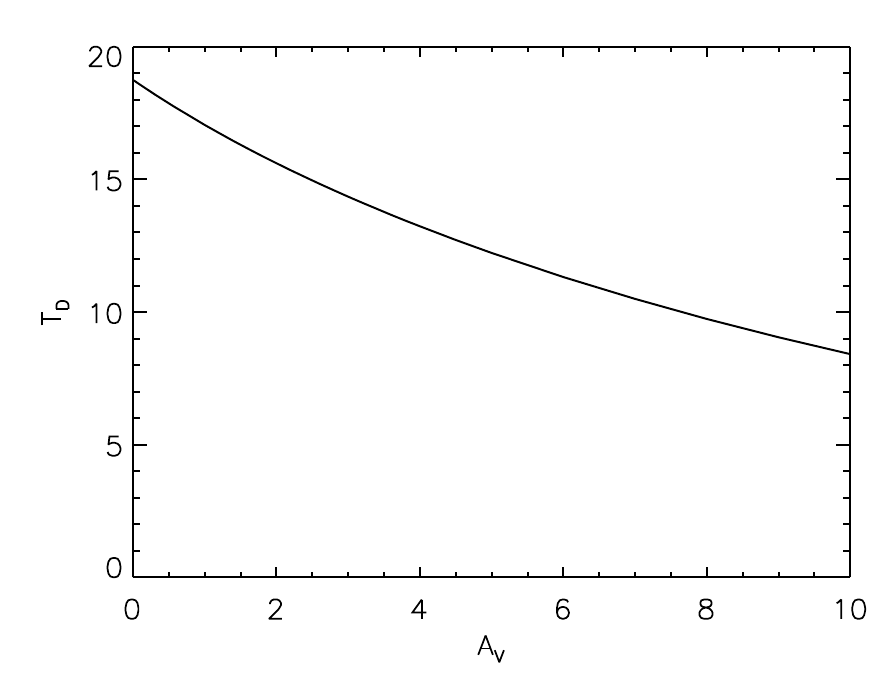}
\caption{Calculated dust temperature through a dark cloud, assuming $a=0.1 \mu$m, as a function of edge-to-center visual extinction.}
\end{figure}

\begin{figure}
\includegraphics[width=0.475\textwidth]{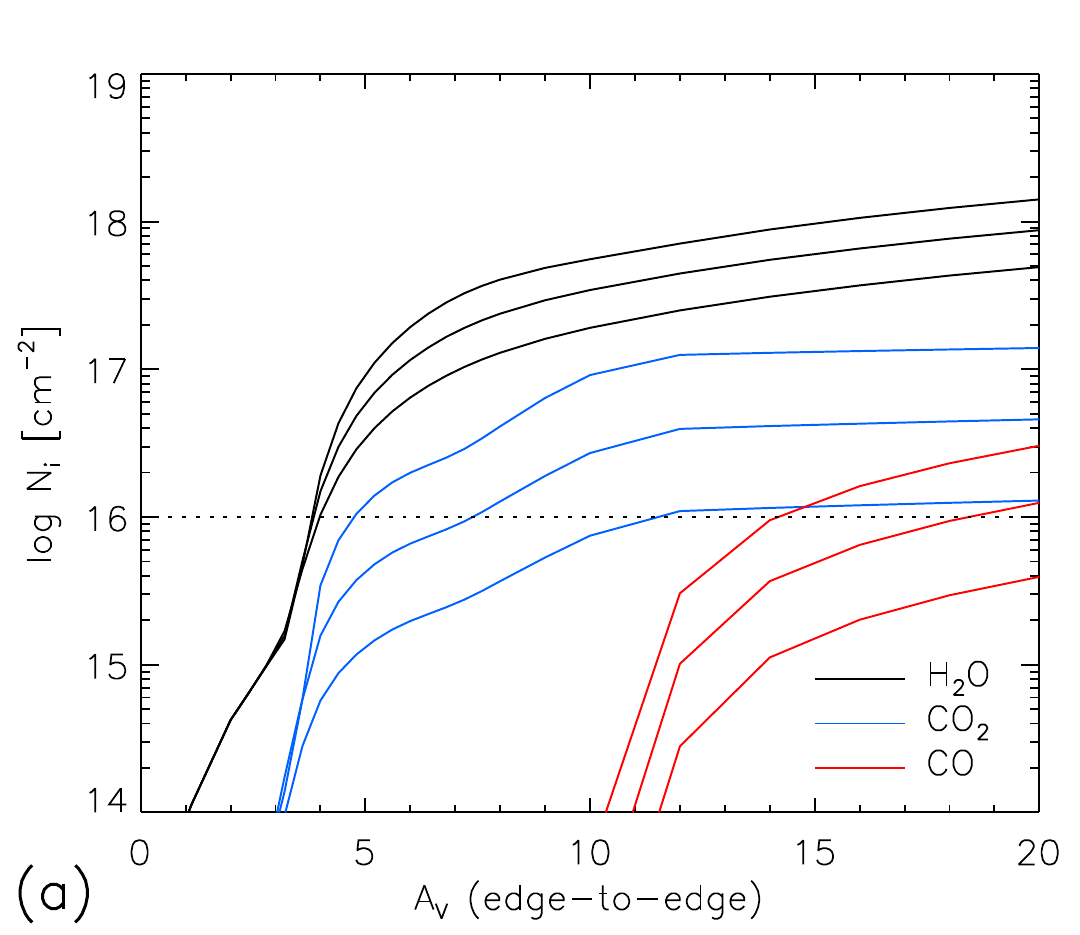}
\includegraphics[width=0.475\textwidth]{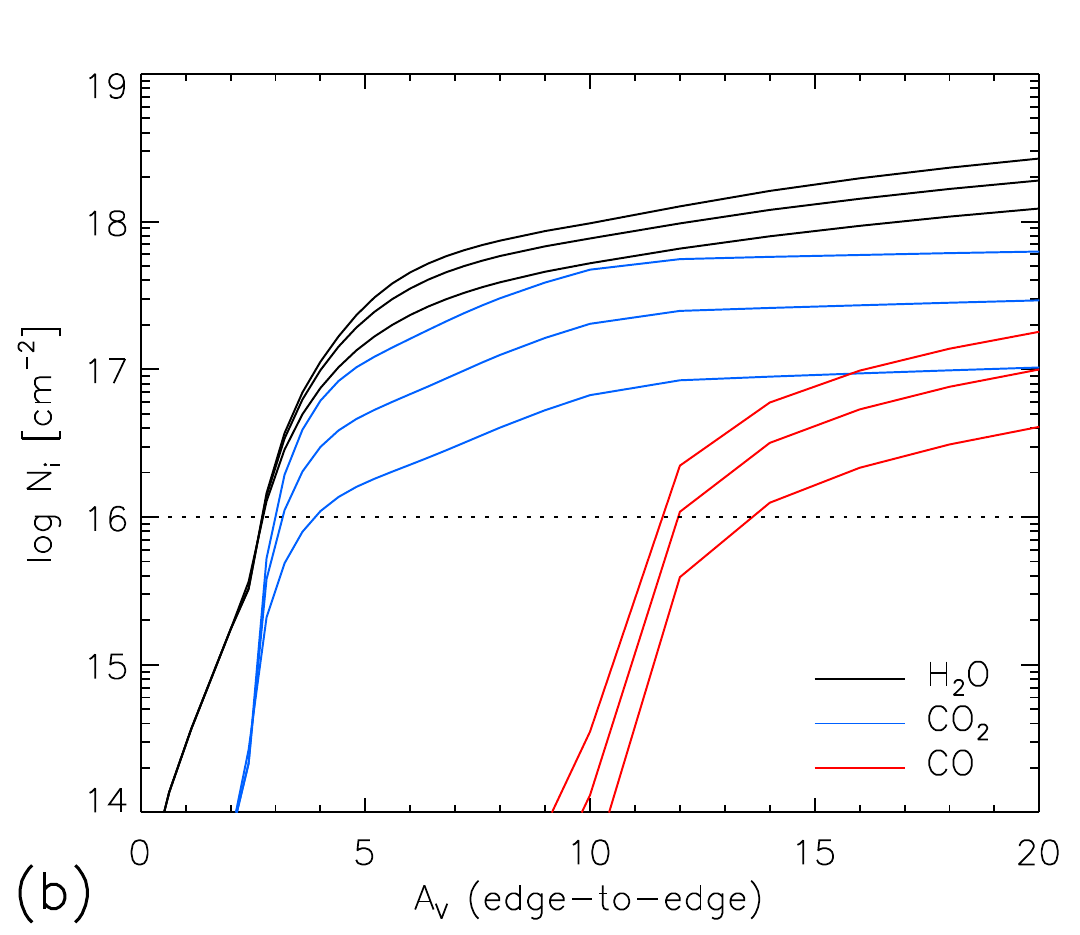}
\includegraphics[width=0.475\textwidth]{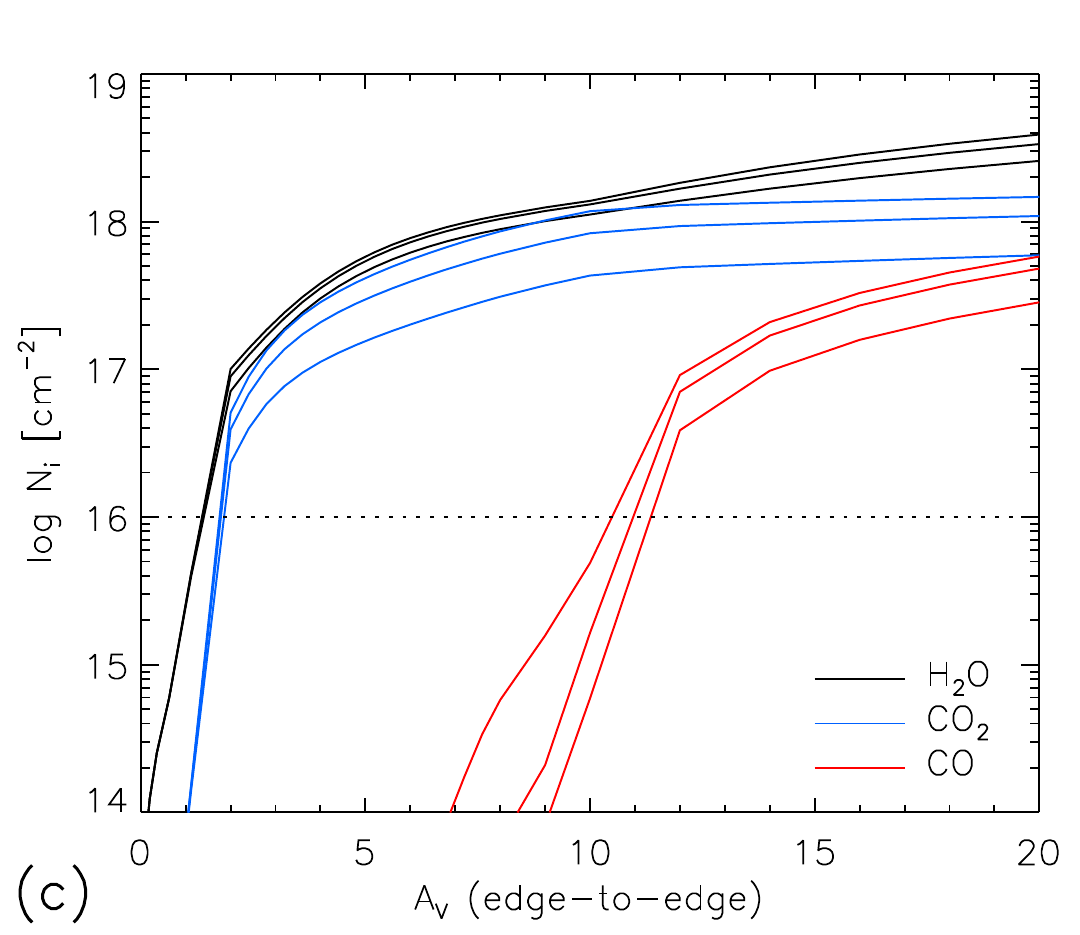}
\caption{Ice column densities integrated through a dark cloud of fixed density, assuming a diffusion barrier height $E_{\mathrm{des}} = 0.3 \, E_{\mathrm{dif}}$. $A_{\mathrm{V}}$ value represents the edge-to-edge visual extinction, i.e. twice the edge-to-center value used within the chemical code. The three curves for each species correspond to timescales of 0.5, 1, and 2 Myr; higher curves represent later times. Each panel shows the model run at a different density: {\bf a)} $n_{\mathrm{H}}=2 \times 10^{3}$ cm$^{-3}$, {\bf b)} $n_{\mathrm{H}}=6 \times 10^{3}$ cm$^{-3}$, {\bf c)} $n_{\mathrm{H}}=2 \times 10^{4}$ cm$^{-3}$}
\end{figure}

\begin{figure}
\includegraphics[width=0.475\textwidth]{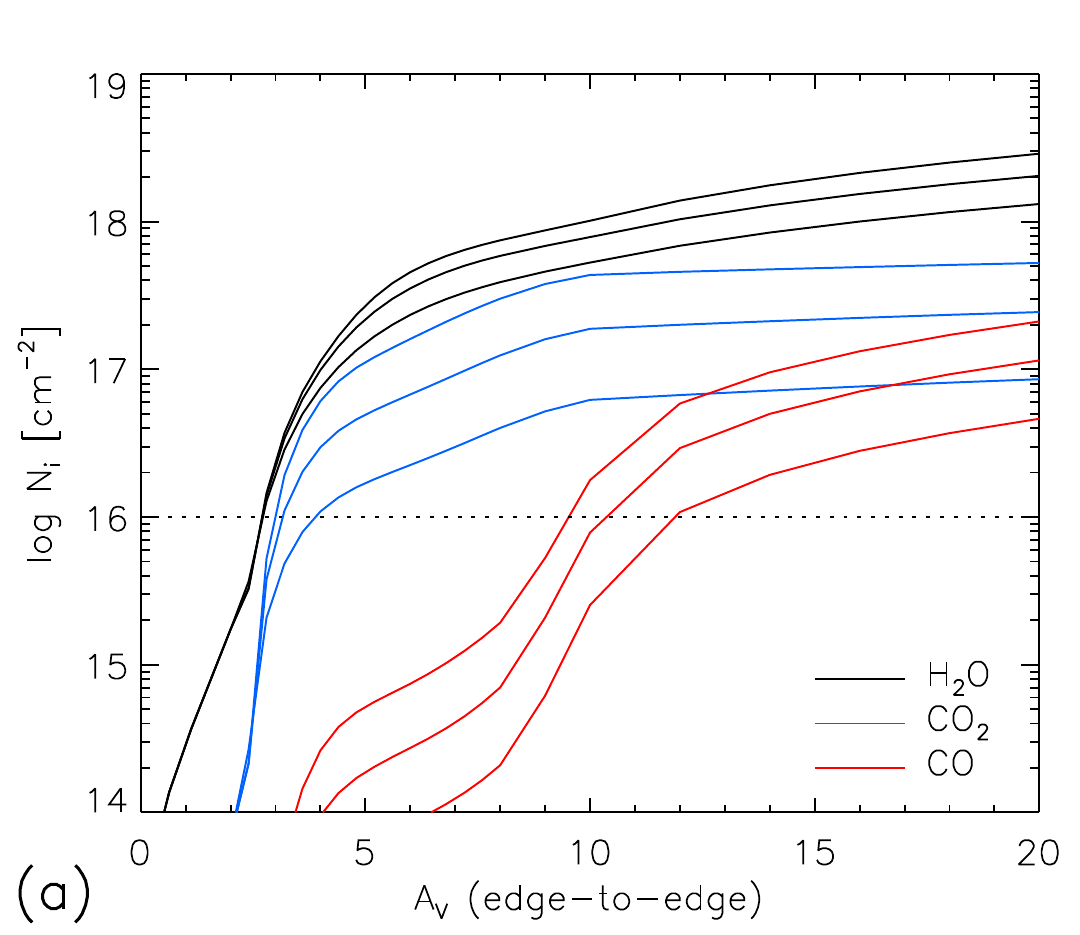}
\includegraphics[width=0.475\textwidth]{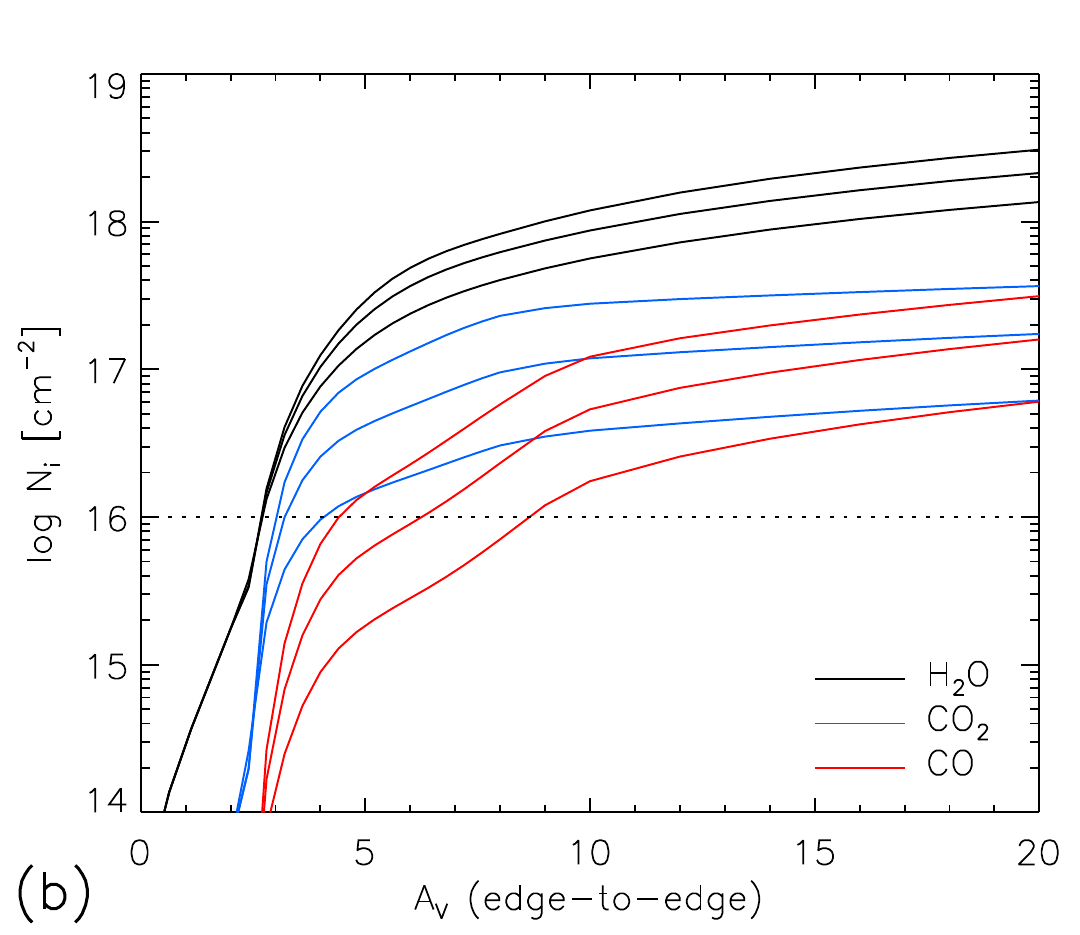}
\includegraphics[width=0.475\textwidth]{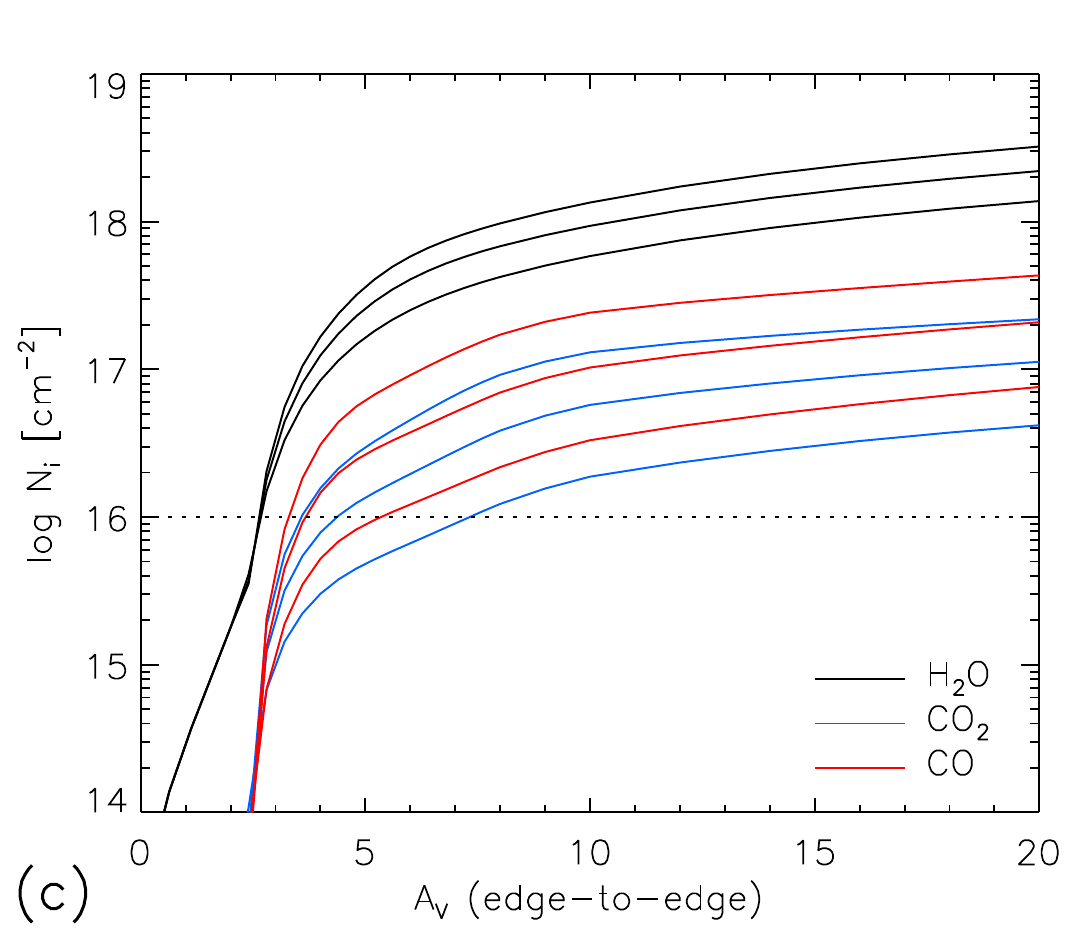}
\includegraphics[width=0.475\textwidth]{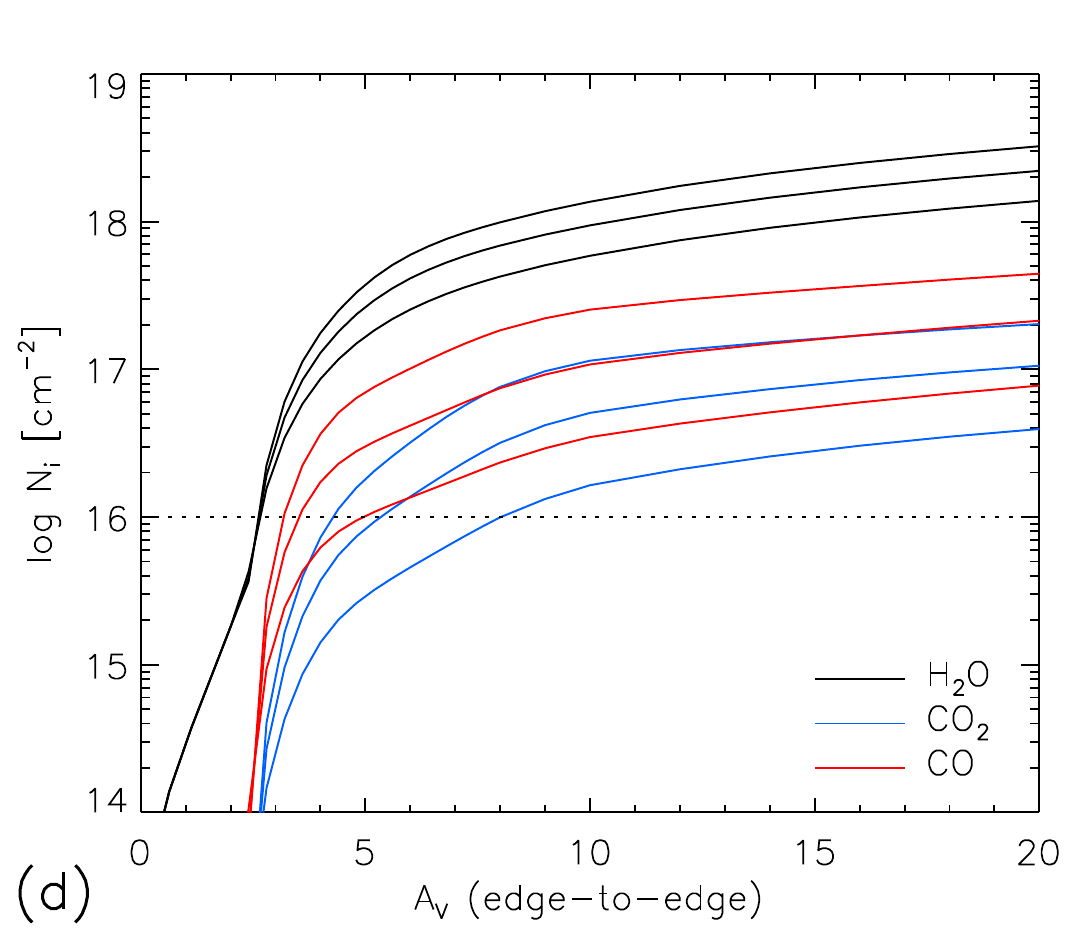}
\caption{Ice column densities integrated through a dark cloud of fixed density $n_{\mathrm{H}}=6 \times 10^{3}$ cm$^{-3}$, assuming a diffusion barrier height {\bf a)} $E_{\mathrm{des}} = 0.35 \, E_{\mathrm{dif}}$; {\bf b)} $E_{\mathrm{des}} = 0.4 \, E_{\mathrm{dif}}$; {\bf c)} $E_{\mathrm{des}} = 0.45 \, E_{\mathrm{dif}}$; {\bf d)} $E_{\mathrm{des}} = 0.5 \, E_{\mathrm{dif}}$. $A_{\mathrm{V}}$ value represents the edge-to-edge visual extinction, i.e. twice the edge-to-center value used within the chemical code. The three curves for each species correspond to timescales of 0.5, 1, and 2 Myr; higher curves represent later times.}
\end{figure}

\begin{figure}
\includegraphics[width=0.475\textwidth]{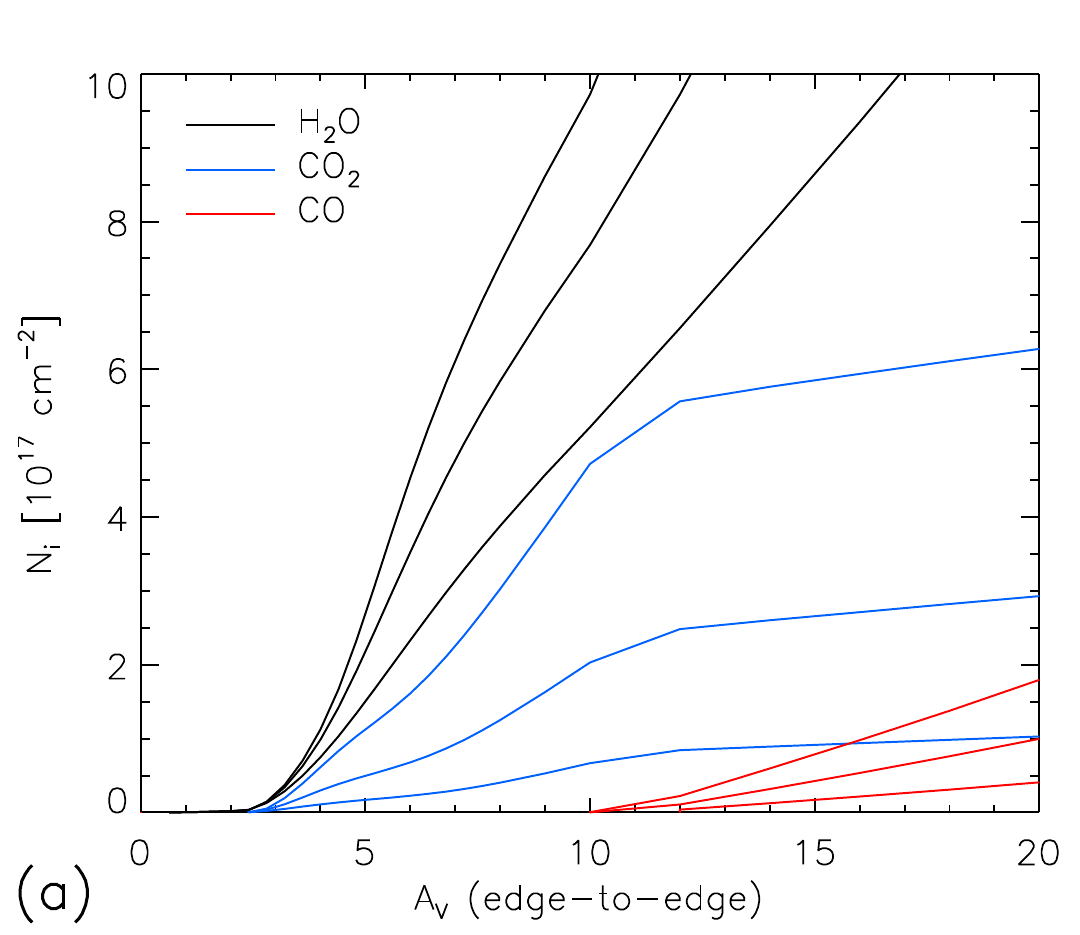}
\includegraphics[width=0.475\textwidth]{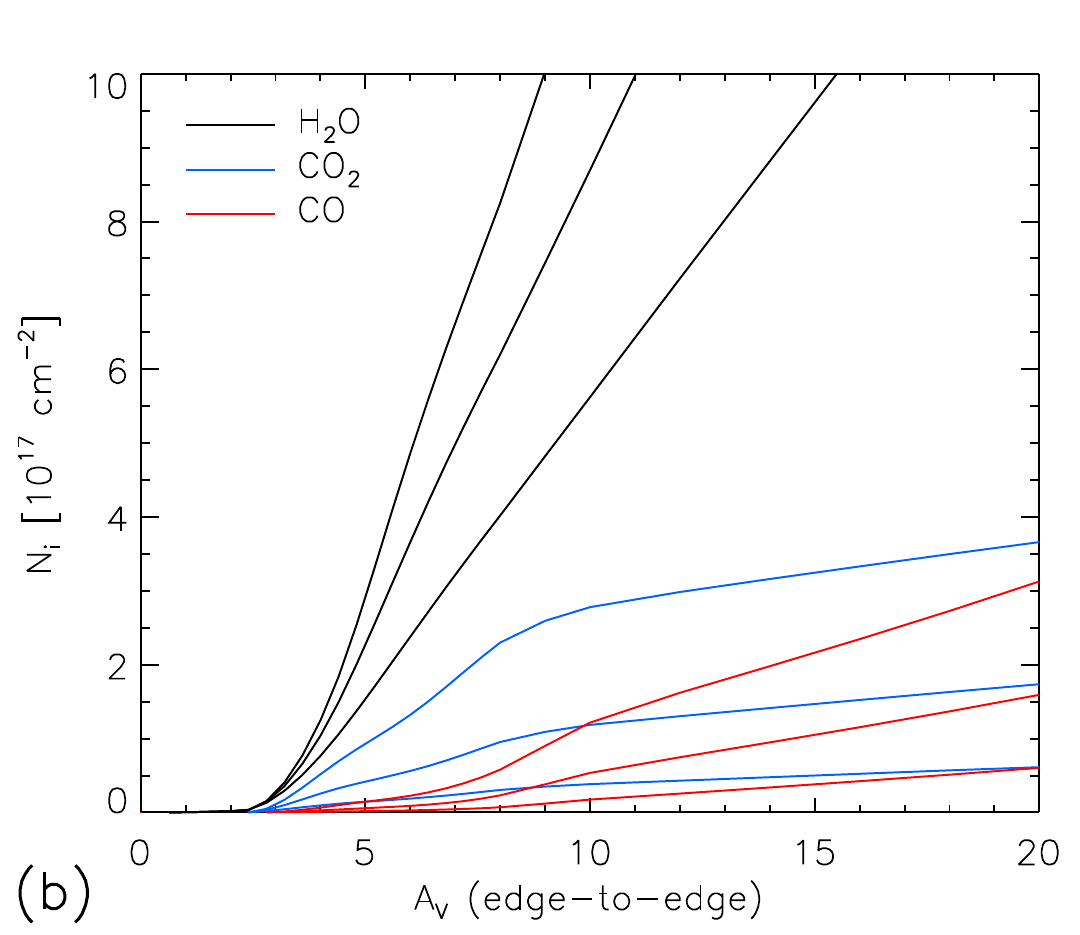}
\caption{Ice column densities integrated through a dark cloud of fixed density $n_{\mathrm{H}}=6 \times 10^{3}$ cm$^{-3}$, assuming a diffusion barrier height {\bf a)} $E_{\mathrm{des}} = 0.3 \, E_{\mathrm{dif}}$; {\bf b)} $E_{\mathrm{des}} = 0.4 \, E_{\mathrm{dif}}$. $A_{\mathrm{V}}$ value represents the edge-to-edge visual extinction, i.e. twice the edge-to-center value used within the chemical code. The three curves for each species correspond to timescales of 0.5, 1, and 2 Myr; higher curves represent later times.}
\end{figure}

\begin{figure}
\includegraphics[width=0.475\textwidth]{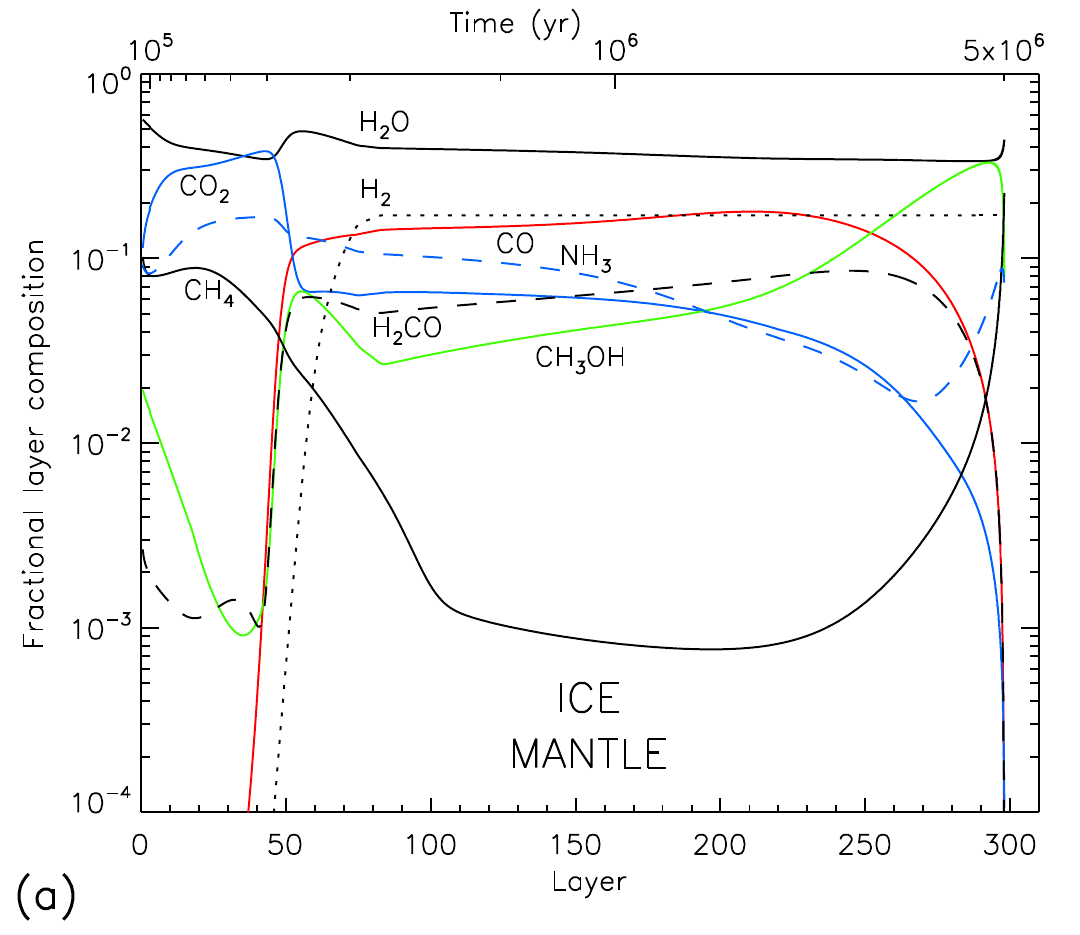}
\includegraphics[width=0.475\textwidth]{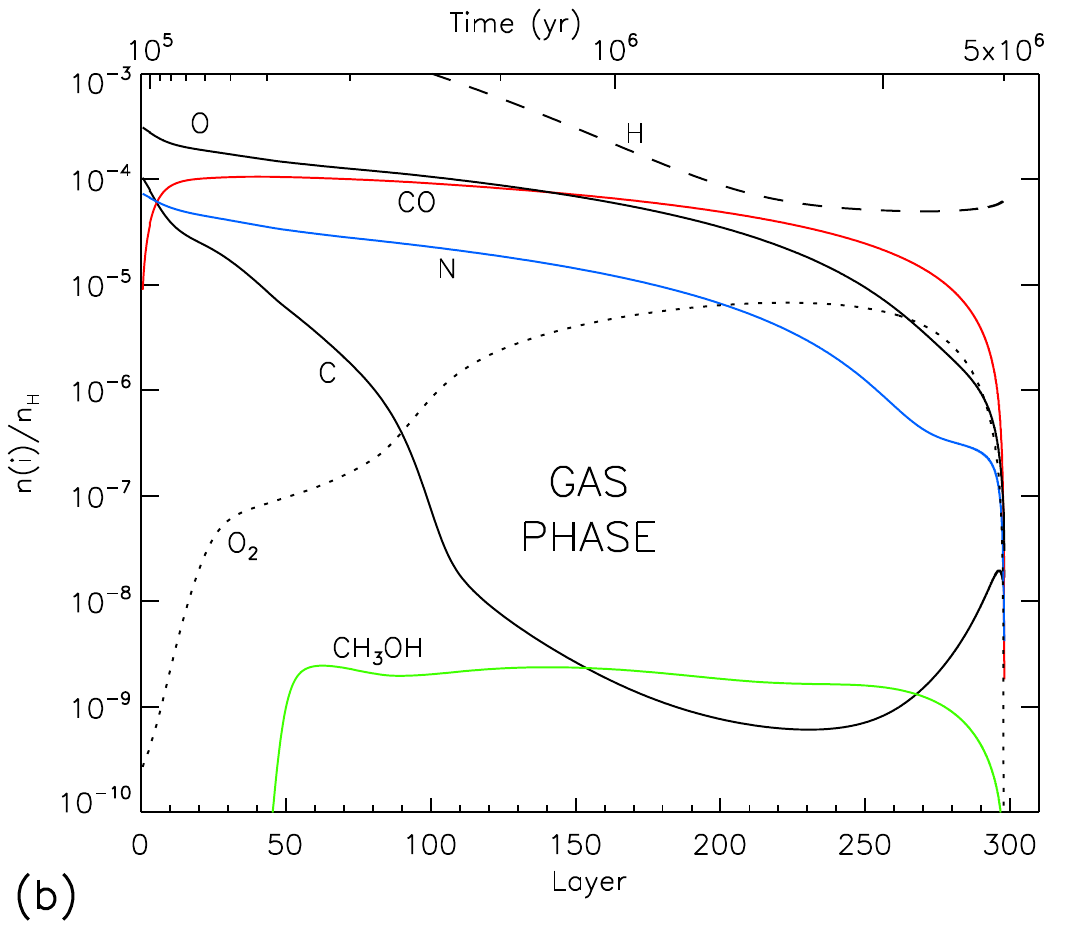}
\includegraphics[width=0.475\textwidth]{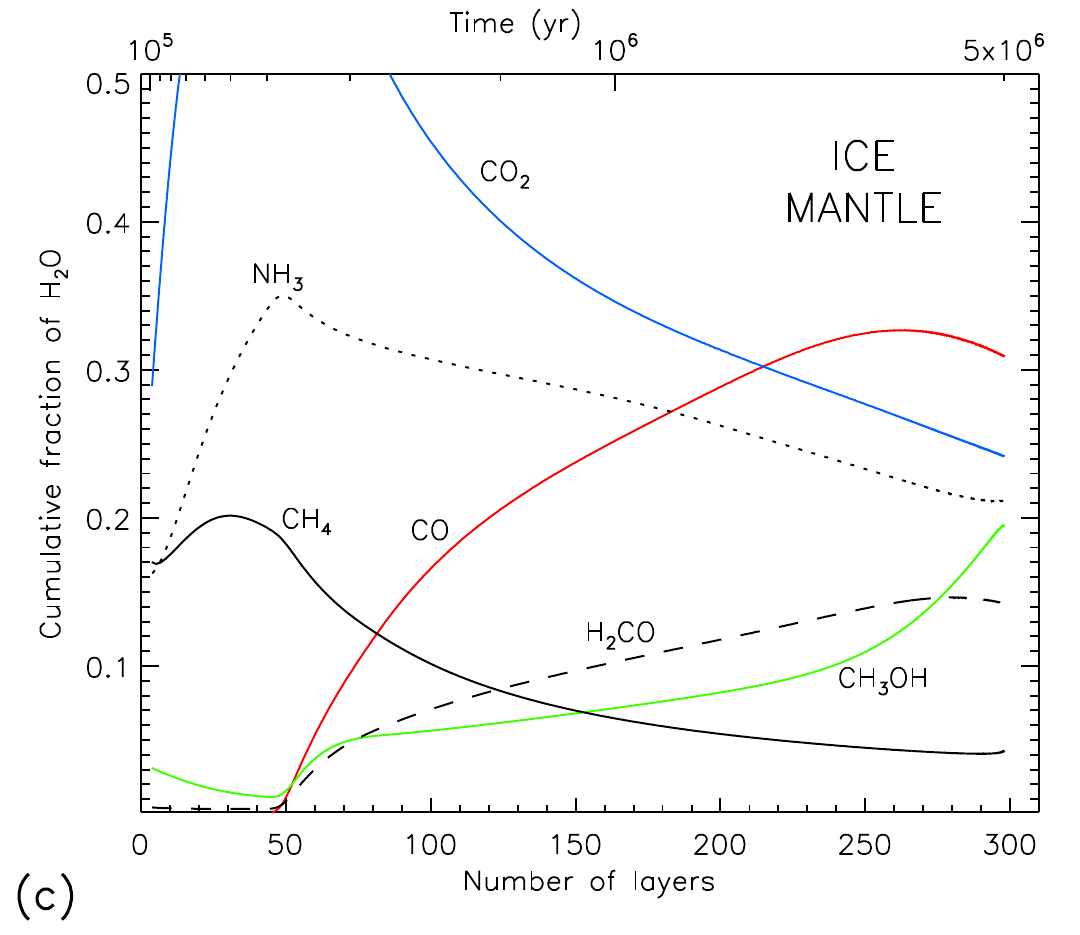}
\caption{{\bf a)} Ice-mantle composition by layer, {\bf b)} gas-phase abundances, and {\bf c)} cumulative fractional composition with respect to water through the ice mantle, for a peak collapse density of $4 \times 10^{4}$ cm$^{-3}$}.
\end{figure}

\begin{figure}
\includegraphics[width=0.475\textwidth]{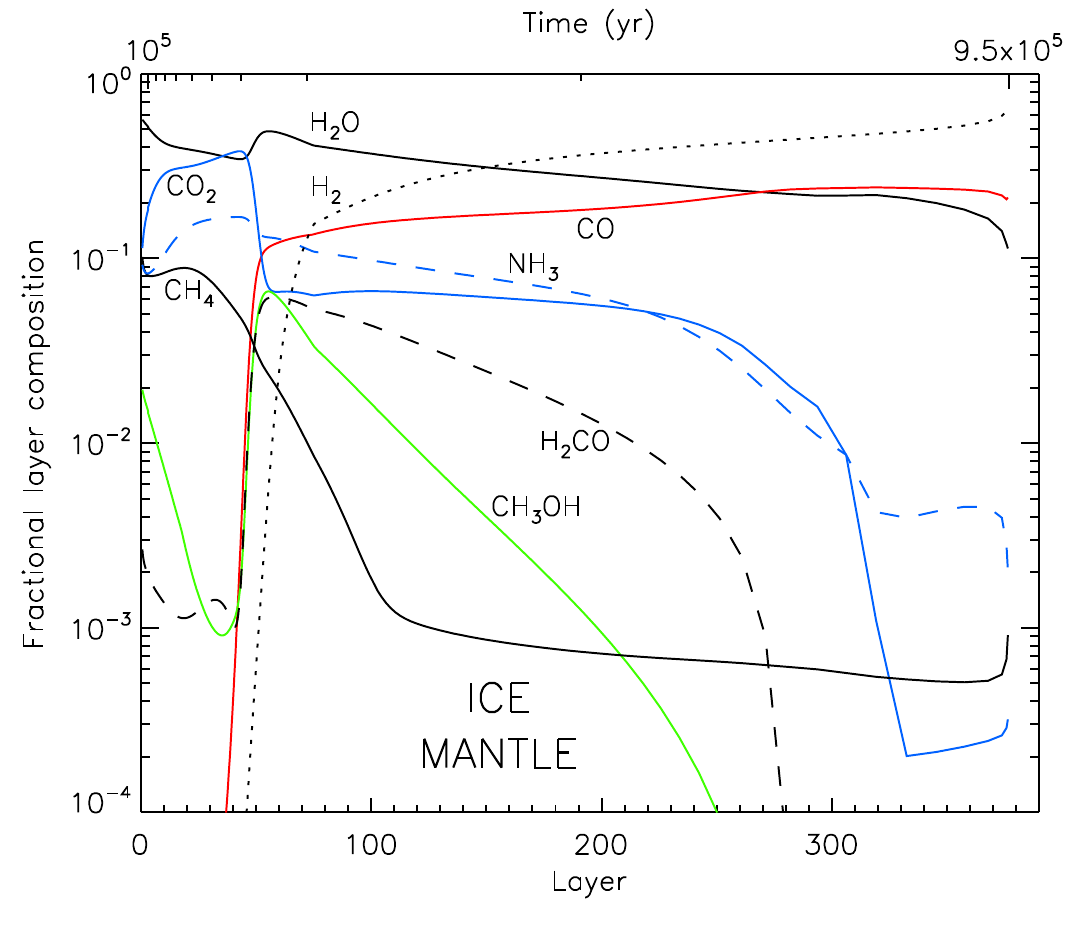}
\caption{Ice-mantle composition by layer, for a peak collapse density of $10^{7}$ cm$^{-3}$}.
\end{figure}

\begin{table}
\begin{center}
\caption{Initial Elemental Abundances}
\begin{tabular}{ll}
\tableline\tableline
Species, $i$ & $n(i) / n_{\mathrm{H}}$$^a$ \\
\tableline
H & 5.0(-5) $^b$ \\
H$_2$ & 0.5 $^b$ \\
He & 0.09 \\
C & 1.4(-4) \\
N & 7.5(-5) \\
O & 3.2(-4) \\
S & 8.0(-8) \\
Na & 2.0(-8) \\
Mg & 7.0(-9) \\
Si & 8.0(-9) \\
P & 3.0(-9) \\
Cl & 4.0(-9) \\
Fe & 3.0(-9) \\
\tableline
\end{tabular}
\tablenotetext{a}{$A(B)=A^{B}$}
\tablenotetext{b}{Static dark-cloud models only}
\end{center}
\end{table}

\begin{table}
\begin{center}
\caption{Binding Energies of Selected Species}
\begin{tabular}{ll}
\tableline\tableline
Species, $i$ & $E_{\mathrm{des}}(i)$ [K] \\
\tableline
H & 450 \\
H$_2$ & 430 \\
C & 800 \\
N & 800 \\
O & 800 \\
CO & 1150 \\
HCO & 1600 \\
OH & 2850 \\
H$_2$O & 5700 \\
\tableline
\end{tabular}
\end{center}
\end{table}

\begin{table}
\begin{center}
\caption{Ice composition, basic dark cloud model$^a$}
\begin{tabular}{llllllll}
\tableline\tableline
Species & Elias 16$^b$ & $1 \times 10^{5}$ yr &  $2 \times 10^{5}$ yr & $5 \times 10^{5}$ yr & $1 \times 10^{6}$ yr &  $2 \times 10^{6}$ yr & Collapse model, $1.2 \times 10^{6}$ yr \\
\tableline
H$_2$O   & 100         &  100  &  100 & 100  &  100 & 100   & 100 \\
CO       & 25          &  5.8  &  10  &  16  &  20  &  20   & 30 \\
CO$_2$   & 18          &  3.3  &  5.7 &  9.2 &  10  &  9    & 30 \\
CH$_4$   & --          &  37   &  30  &  19  &  14  &  12   & 5.1 \\
CH$_3$OH & $< 3$       &  0.34 &  1.1 & 4.4  &  9.7  &  18  & 8.7 \\
H$_2$CO  & --          &  1.2  &  2.8 & 6.4  &  9.3  &  10  & 12 \\
NH$_3$   & $ \leq 9 $    &  25   &  25  &  24  &  22  &  19 & 25 \\
\tableline

\tableline
\end{tabular}
\tablenotetext{a}{Observational values for Elias 16 as collated by Gibb et al. (2000)}

\end{center}
\end{table}

\end{document}